\pdfoutput=1
\documentclass[%
 reprint,
superscriptaddress,
a4paper,
longbibliography,
amsmath,amssymb,
aps,
numbers,
prb,
ulem
]{revtex4-1}

\usepackage[pdftex]{hyperref,graphicx}
\usepackage[usenames, svgnames]{xcolor}
\usepackage{amsfonts,amssymb,amsmath,mathrsfs}
\usepackage[separate-uncertainty=true]{siunitx}
\usepackage[english]{babel}
\usepackage[utf8]{inputenc}
\usepackage[T1]{fontenc}
\usepackage{color,soul}
\usepackage{xspace}
\usepackage{upgreek}
\usepackage[normalem]{ulem}
\usepackage{comment}
\usepackage{xfrac}

\usepackage[caption=false]{subfig}
\usepackage{amsthm, amssymb}
\usepackage{ulem}

\newcommand{\ket}[1]{|#1\rangle}
\newcommand{\bra}[1]{\langle#1|}
\newcommand{\fket}[1]{| #1 \rangle \rangle}

\newcommand{\tket}[1]{| #1 )}

\newcommand{\figref}[2]{\hyperref[#1]{\getrefnumber{#1}(#2)}}
\hypersetup{
 pdftitle = {Dissipation engineered direction-dependent filter for quantum ratchets},
 pdfauthor = {Zlata Fedorova, Christoph Dauer, Sebastian Eggert, Johann Kroha, Stefan Linden},
	colorlinks=true,linkcolor=blue,citecolor=blue,filecolor=blue,urlcolor=blue,pdfpagemode=UseNone,pdfstartview={XYZ null null 1.00}
}

\addto\captionsenglish{}
\addto\captionsenglish{}

\renewcommand\textemdash{\leavevmode\unskip\kern0.8pt\rule[0.19\baselineskip]{8pt}{0.4pt}\kern1pt\ignorespaces}

\begin{document}

\preprint{APS/123-QED}

\title{Dissipation engineered directional  filter for quantum ratchets}

\author{Zlata Fedorova}
\thanks{Equal contribution}
\email{cherpakova@physik.uni-bonn.de}
\affiliation{Physikalisches Institut, Rheinische Friedrich-Wilhelms-Universit\"at Bonn, Nussallee 12, 53115 Bonn, Germany.}

\author{Christoph Dauer}
\thanks{Equal contribution}
\email{cdauer@physik.uni-kl.de}
\affiliation{Physics Department and Research Center OPTIMAS, TU Kaiserslautern, 67663 Kaiserslautern, Germany.}

\author{Anna Sidorenko}
\email{a_sidorenko@uni-bonn.de}
\affiliation{Physikalisches Institut, Rheinische Friedrich-Wilhelms-Universit\"at Bonn, Nussallee 12, 53115 Bonn, Germany.}

\author{Sebastian Eggert}
\email{eggert@physik.uni-kl.de}
\affiliation{Physics Department and Research Center OPTIMAS, TU Kaiserslautern, 67663 Kaiserslautern, Germany.}

\author{Johann Kroha} 
\email{kroha@th.physik.uni-bonn.de}
\affiliation{Physikalisches Institut, Rheinische Friedrich-Wilhelms-Universit\"at Bonn, Nussallee 12, 53115 Bonn, Germany.}

\author{Stefan Linden}
\email{linden@physik.uni-bonn.de}
\affiliation{Physikalisches Institut, Rheinische Friedrich-Wilhelms-Universit\"at Bonn, Nussallee 12, 53115 Bonn, Germany.}

\date{\today}

\begin{abstract}
We demonstrate transport rectification in a hermitian Hamiltonian quantum ratchet by a dissipative, dynamic impurity.
While the bulk of the ratchet supports transport in both directions, the properly designed loss function of the local impurity acts as a direction-dependent filter for the moving states. We analyse this scheme theoretically by making use of Floquet-S-Matrix theory. In addition, we provide the direct experimental observation of  one-way transmittance in periodically modulated plasmonic waveguide arrays containing a local impurity with engineered losses.
\end{abstract}

\pacs{Valid PACS appear here}
\maketitle

\section{Introduction}

Systems governed by time-periodic Hamiltonians can feature a variety of novel transport phenomena inaccessible in equilibrium. A fascinating example is the ratchet effect, i.e. the ability to convert periodic drive into directed motion without a bias force. The working principle of a ratchet relies on the breaking of space- and time-reversal symmetry which would otherwise not allow a directed current~\cite{denisov2014tunable}. Introduced by Smoluchowski~\cite{smoluchowski1927experimentell} and Feynman~\cite{feynman1963feynman}, ratchets represent a wide class of microscopic motors, which operate in classical as well as in quantum systems. In particular, the ratchet effect was observed in microbiological~\cite{mahmud2009directing} and molecular motion~\cite{serreli2007molecular}, semiconductor~\cite{linke1999experimental} and superconductor~\cite{costache2010experimental} heterostructures, irradiated graphene~\cite{drexler2013magnetic}, electron pumps~\cite{lehmann2002molecular}, photonic setups~\cite{zhang2015experimental,dreisow2013spatial}, and Bose-Einstein condensates~\cite{salger2009directed,ni2017hamiltonian}. 

Most classical ratchets are based on thermal motion and dissipation where initial conditions play no role~\cite{serreli2007molecular,mahmud2009directing,hanggi2009artificial}. In contrast, in quantum ratchets directed transport arises from a quantum coherence effect, namely the Chern number or Berry phase accumulated when a quantum state is moved by the driving potential along a closed loop in Hamiltonian parameter space. When the driving is adiabatic, the transport current is quantized, known as Thouless pumping \cite{thouless1983quantization}. The realization of this concept faces, however, two fundamental difficulties. For fast, non-adiabatic driving the transport quantization generically breaks down, and the transport efficiency depends sensitively on the relative phase of the driving parameters and on the initial state of the driven system~\cite{denisov2007periodically,gong2007dissipationless,salger2009directed,ni2017hamiltonian,fedorova2020observation}. 
More concretely, being periodic in time, quantum ratchets can be described in terms of  Floquet states \cite{denisov2007periodically}. The overlap of these states with the initial conditions determines their population and hence their contribution to the  current in the stationary state. Since the system's Hamiltonian can support currents in both directions, only a proper choice of the initial conditions will generate maximal, unidirectional transport. In contrast, it is desirable to achieve optimal transport efficiency without initial-state preparation. 

\begin{figure}[t] 
\includegraphics[width=\linewidth]{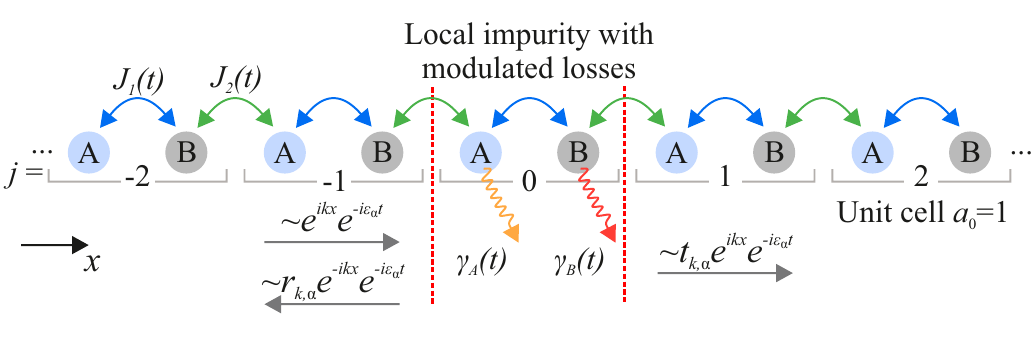}
\centering
\caption{Sketch of the dimerized tight-binding Floquet chain with local time-periodically modulated decay rates $\gamma_{{\rm A},\rm B}(t)$ and hopping amplitudes $J_1(t)$ and $J_2(t)$. We partition the chain in sublattices A/B and label the unit cell by $j$. The reflection $r_{k,\alpha}$ and transmission $t_{k,\alpha}$ of the Floquet state with quasimomentum $k$ and band index $\alpha$ are schematically indicated by arrows.}
\label{fig:Modelintro}
\end{figure}

In the present study we 
propose and experimentally realize a scheme for quantized, directional transport in fast Hamiltonian ratchets using a local impurity with engineered dynamic dissipation as a direction-dependent filter. In the driven Rice-Mele model where space- and time-inversion symmetries
are broken by the driven Hamiltonian,
any initial condition can carry a current. However, the topological transport quantization is not robust to nonadiabatic effects~\cite{lohse2016thouless,nakajima2016topological} unless intensity or particle losses are introduced globally~\cite{fedorova2020observation,fehske2020quantized}. Since adiabatic conditions cannot be reached in most experimental situations and it is often desirable to minimize losses, we here consider the periodically driven Su-Schrieffer-Heeger (SSH) model. In general it breaks time-inversion symmetry due to a phase shift between the time-periodic coupling constants, but always preserves space-inversion on the Hamiltonian level~\cite{su1979solitons,dreisow2013spatial,kartashov2016diffraction,budich2017helical}. As we will show below, this system supports quantized transport for certain, non-adiabatic driving frequencies once the space inversion symmetry is broken by initial conditions.
 This model as well as its transport properties are discussed in Section~\ref{sec:Ratchet model}. In Section~\ref{sec:Direction-dependent filter} we introduce time-dependent losses localized at a finite number of lattice sites (see Fig.~\ref{fig:Modelintro}).
By means of the Floquet formalism, we show how a properly designed, time-periodic local loss function can facilitate non-reciprocal transport through this impurity. In previous studies, local periodic driving of the real part of a potential has been used to control transmission through the modulated region~\cite{thuberg2017perfect,reyes2017transport,agarwala2017effects}.
 In this work the non-Hermiticity of the impurity is a key feature, as it breaks the relevant space and time inversion symmetries in the scattering process and thus enables the non-reciprocal transport \cite{Moskalets2002,Li2018,fedorova2020observation}. In Section~\ref{sec:Floquet S-Matrix analysis} we develop a numerical method based on the Floquet-S-Matrix theory \cite{Smith2015,Millack1990} in order to analyse the direction-dependent transmission coefficients in dependence on the system parameters. We find the optimal driving scheme to achieve the largest asymmetry in the transmission for a given decay rate. 
Furthermore, in Section~\ref{sec:Experiments} we provide the experimental observation of transport rectification in arrays of coupled dielectric-loaded surface plasmon-polariton waveguides (DLSPPW) with controlled losses. A brief summary and concluding remarks are given in Section~\ref{sec:Conclusion}.

\begin{figure}[t] 
\includegraphics[width=\linewidth]{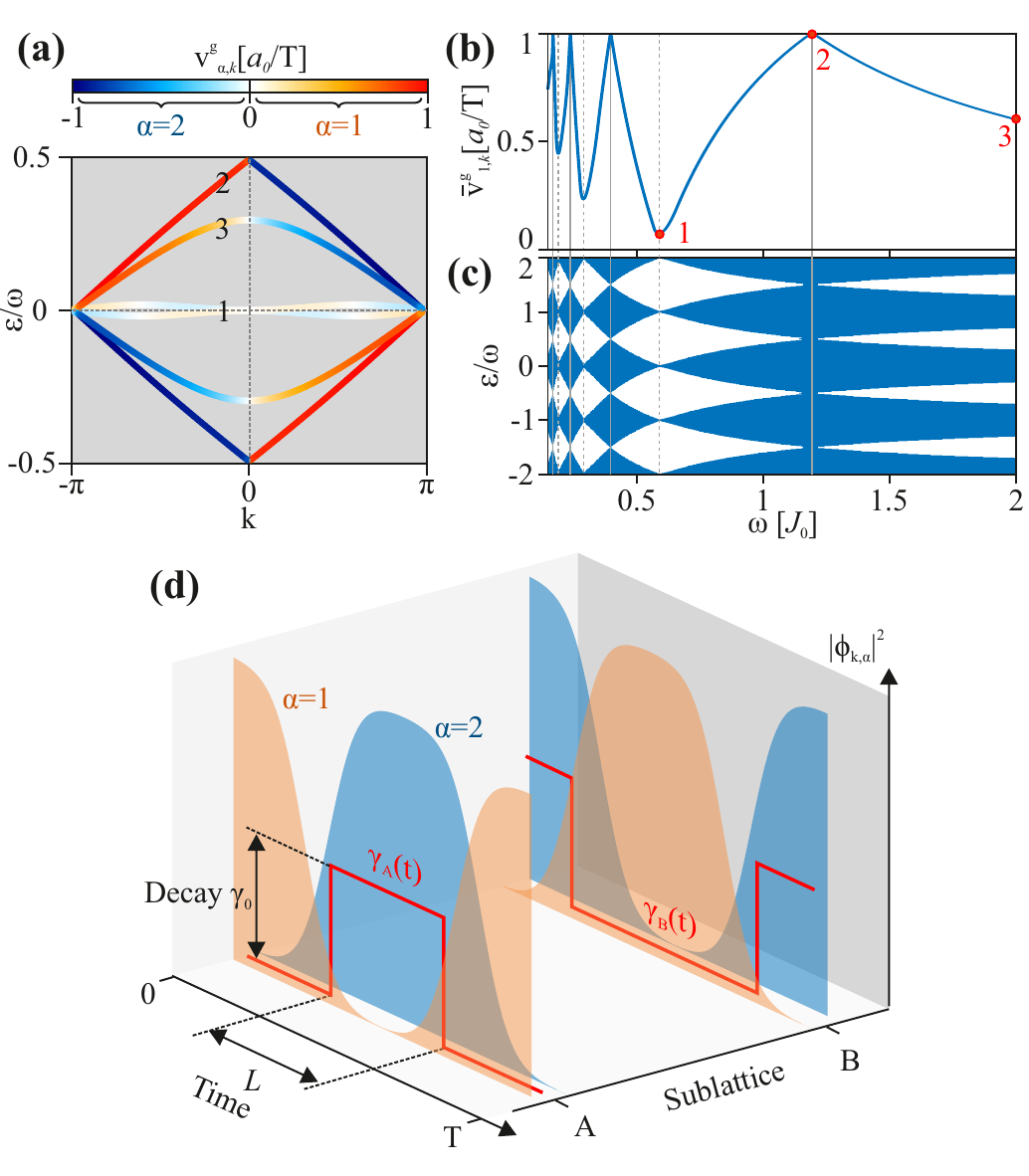}
\centering
\caption{(a)~Band structure  in the $1^{\mathrm{st}}$ FBBZ at three different driving frequencies: 1: $\omega=0.597$, 2: $\omega=1.195 J_0$, 3: $\omega=2 J_0$. The color-code shows the corresponding group velocity $v^\mathrm{g}_{\alpha,k}$. (b)~Average absolute value of the group velocity as a function of $\omega$.  Grey lines mark values from Eq.~(\ref{eq:DissConditionLinearBands})-(\ref{eq:DissConditionFlatBands}) and numbers highlight the frequencies from (a). (c)~Quasienergy spectra in dependence on the driving frequency. (d)~Squared absolute value of the state amplitude at $\omega=1.195J_0$ with  positive ($\alpha=1$) and negative ($\alpha=2$) group velocity along one period at  sublattices A and B. Red lines show the time-dependent losses $\gamma_{A/B} (t)$ on sublattices A/B for $\varphi=0$ which consist of
temporal intervals $L$ of the constant decay rate $\gamma_0$. Note, that in our model $\gamma_{A/B} (t)$ are applied only to the central unit cell as illustrated in Fig.~\ref{fig:Modelintro}.}
\label{fig:Model}
\end{figure}

\section{Ratchet model}\label{sec:Ratchet model}
 
 The SSH model~\cite{su1979solitons} consists of a dimerized tight-binding chain with a two-site unit cell and constant, homogeneous onsite potentials (sublatice A: odd sites, sublattice B: even sites; see Fig.~\ref{fig:Modelintro}). Its Hamiltonian is given by
\begin{equation}
\label{eq:BulkHamiltonian}
H_{\rm Bulk}(t)=\sum_j J_1(t) c_{j,\rm A}^\dagger c^{\phantom{\dagger}}_{j,\rm B}+J_2(t) c_{j,\rm B}^\dagger c^{\phantom{\dagger}}_{j+1,\rm A}+h.c.~.
\end{equation}
Here, $c_{j,\rm{A}/\rm B}^{\dagger}$/$c_{j,\rm{A}/\rm B}$ are creation/annihilation operators for site $A/B$ in unit cell $j$. The time-periodic intra-/intercell coupling $J_{1/2}(t)$ constants are modulated such that $J_1(t) > J_2(t)$ holds for the first half of the period while the situation is inverted in the second half. In this case the movement of a right-moving excitation from $A$ to $B$ sites in the first half of the period and then from $B$ to $A$ sites can be supported at the special group velocity of one unit cell per driving period and we find that the transport may become largely independent of the magnitude of the quasimomentum. 
In the experiments below we sinusoidally vary the spacing between the neighboring sites, which leads to the following functional form of the coupling constants
\begin{subequations}
\label{eq:Hopping}
\begin{align}
    &J_{1}(t)=J_0 e^{-\lambda(1-\sin(\omega t))},\\ &J_{2}(t)=J_1(t-T/2).
\end{align}
\end{subequations}
Here, $\omega=2\pi/T$ is the driving frequency and $T$ is the period of modulation.
To be consistent with the experiment, we choose in the following $\lambda=2.11$. For simplicity, we measure all the quantities  in units of $J_0$ and set the unit cell as well as the reduced Planck constant  to one: $a_0=1$ and $\hbar=1$. The calculation of the bulk quasienergy spectrum is carried out using the Floquet-Bloch  theory~\cite{gomez2013floquet} (see Appendix~\ref{appendix A}) and exemplified in Fig.~\ref{fig:Model}~(a) in the $1^{\mathrm{st}}$ Floquet-Bloch Brillouin zone (FBBZ) for three different $\omega$ values. The figure shows that the driving frequency has a huge impact on the band shape:  At $\omega=1.195J_0$ (num.2) the bands are  almost linear with the slope $1/T$, in contrast, at $\omega=0.597$ (num.1) they become almost flat. Note that the almost linear, gapless bands are helical in the Floquet-Bloch Brilloin zone and can be related to a non-trivial topology \cite{budich2017helical}. The chiral symmetry of the Hamiltonian~(\ref{eq:BulkHamiltonian}) guarantees that the spectrum is always symmetric with respect to the Floquet quasienergy $\epsilon\to -\epsilon$. We can, therefore, choose to label the quasienergies and the corresponding Floquet-Bloch states according to the sign of the group velocity such that $\alpha=1$ stands for $v^\mathrm{g}\geq 0$ while $\alpha=2$ for $v^\mathrm{g}\leq 0$. As shown in Fig.~\ref{fig:Model}~(b), the group velocity averaged over all ($\alpha=1$)-states, $\frac{1}{2\pi}\int_{-\pi}^{\pi} dk v^\mathrm{g}_{1,k}$, depends oscillatory on $\omega$. The quasienergy spectrum in Fig.~\ref{fig:Model} (c) reveals that such a behavior is directly connected to the oscillating size of the band gap.  The ratchet transport is most efficient when the average group velocity reaches its maximum of one unit cell per driving period. These points correspond to gapless helical bands, which in turn possess minimal dispersion.
Such a dynamics can be qualitatively understood with a simplified dimer model which is discussed in Appendix~\ref{appendix B}. There we find that the helical bands and maximum group velocity occur if the states undergo half-cycles of Rabi oscillations  between the two sublattices. This physics can be linked to the condition for the velocity maxima and gap closings in terms of the time-integrated hopping
\begin{equation}
\label{eq:DissConditionLinearBands}
    \omega_n=\frac{4\int_0^{1}d\xi J_{i}(2 \pi /\omega~\xi)}{1+2 n},\quad n\in \mathbb{N}_0,~i=1,2.
\end{equation}
Likewise, minima of the group velocity and the bandwidth are predicted to occur near 
\begin{equation}
\label{eq:DissConditionFlatBands}
    \omega'_n=\frac{4\int_0^{1}d\xi J_{i}(2 \pi /\omega~\xi)}{2+2 n},\quad n\in \mathbb{N}_0,~i=1,2,
\end{equation}
where the transport is partially blocked. 
In Figs.~\ref{fig:Model} (b,c) it is shown that this correctly predicts extrema of the average group velocity and the closing of the bulk gap in our system.
In the context of our experimental setup, the achievable frequency range is roughly $\omega\in [0.7J_0; 2J_0]$. Further, we will focus only on the first maximum $\omega_{0}=1.195J_0$ as it lies in this range. 
The dynamics of the periodic Floquet-Bloch states in this regime is governed by oscillations between the two sublattices [see Fig.~\ref{fig:Model}~(d)]: The density of a right moving state ($\alpha=1$) tunnels from sublattice A at $t=Tm$ to sublattice $B$ at $t=T/2+Tm$, while it is fully transferred back at $t=T(m+1)$, where $m$ is an integer. Time-reversal symmetry is present in our model due to the special phase relation between the time periodic coupling constants \eqref{eq:Hopping}. This symmetry implies that the aforementioned process is exactly inverted for the left-moving state with the same quasienergy. The ratchet effect occurs when the asymmetric initial conditions are applied at t=0, i.e.~the system is initiated at one sublattice A or B. Via Fourier analysis, this results in predominant population of only the left- or right- moving states, respectively. Thus, such an SSH-based ratchet strongly depends on the initial conditions, in contrast to Thouless pumping, where transport is induced by the phase relation between the on-site energies and the coupling constants.


\section{Direction-dependent filter}\label{sec:Direction-dependent filter}

Now, we additionally subject two central lattice sites, A and B from unit cell $j=0$, to time-periodic losses oscillating at the same frequency $\omega$ as the bulk (see Fig.~\ref{fig:Modelintro}). 
The corresponding Hamiltonian is the sum of the bulk Hamiltonian and the local impurity $V(t)$:
\begin{subequations}
\label{eq:ModelHamiltonian}
\begin{align}
&H(t)=H_{\rm Bulk}(t)+V(t),
	\\
	\label{eq:ModelPotential}
&V(t)=-i\gamma_{\rm A}(t) c_{0,\rm A}^\dagger c^{\phantom{\dagger}}_{0,\rm B}-i\gamma_{\rm B}(t) c_{0,\rm B}^\dagger c_{0,\rm B}^{\phantom{\dagger}}.
	\end{align}
\end{subequations}
The decay rates on sublattices $A/B$ are denoted by $\gamma_{{\rm A}/\rm B}(t)$ and have a form of $T$-periodic step functions, i.e. the onsite losses can be turned on and off in a periodic manner as realized in our experiments below.
The mathematical description is given by
\begin{subequations}
 \label{eq:LossFunction}
    \begin{align}
       &\gamma_{\rm A}(t)=\gamma_0\cdot\Theta\boldsymbol{(}-\cos(\omega t+\varphi)-\cos(\pi L/T)\boldsymbol{)},\\ 
       &\gamma_{\rm B}(t)=\gamma_{\rm A}(t-T/2),  
    \end{align}
\end{subequations}

 where $\Theta(x)$ is the Heaviside function, $\gamma_0$ the loss amplitude, $L$ the duration of the losses within one period $T$ ($L<T$), and $\varphi$ is the phase shift. Note, that for $\varphi=0$ the losses are out of phase with the coupling constants, i.e. they are centered at $t=(m+1/2)~T$ on sublattice A and at $t=m~T$ on sublattice B.  
 
 In the following we aim at analyzing the scattering process of a quantum particle propagating along the driven SSH chain and scattered by this impurity (see gray arrows in Fig.~\ref{fig:Modelintro}).
 Assume that the system is driven with the resonant frequency $\omega_0$, so that the bands are helical and the band-averaged group velocity is maximal.
 In order to understand the origin of non-reciprocal transmission induced by $V(t)$, it is central to look at the periodic exchange of the state density between the two sublattices. In the previous section it was shown that the counter-propagating Floquet-Bloch states have different spatio-temporal distributions which enabled to populate only states moving in the chosen direction by proper choice of the initial conditions. The same feature can be employed for direction-dependent filtering. In particular, introducing strong losses at space-time moments, where the maxima of $|\phi_{k,\alpha=2,m}(t)|^{2}$ reside, the  time-reversal symmetry and the oscillatory motion of the states guarantees that $|\phi_{k,\alpha=1,m}(t)|^{2}$ is minimal at these moments [see red \textcolor{violet}{lines} in Fig.~\ref{fig:Model}~(d)]. Thus, we can effectively absorb only the states moving in $-x$ direction ($\alpha=2$~states). The transmission through the impurity can be controlled by tuning such system parameters as $\gamma_0$, $L$, and $\varphi$.  It is to be expected that with increasing loss strength $\gamma_0$ and duration of the losses $L$ transmission in $-x$ direction decreases.  But how strongly will the states moving in the opposite direction be affected? What is the influence of the phase shift $\varphi$? Does one-way transmission persist at frequencies away from resonance? To answer these questions and to predict optimal parameters for the experiment we apply Floquet S-matrix theory.

\section{Floquet S-Matrix analysis}\label{sec:Floquet S-Matrix analysis}
\begin{figure}[tbtbtbt]
\centering
\includegraphics[width=\linewidth]{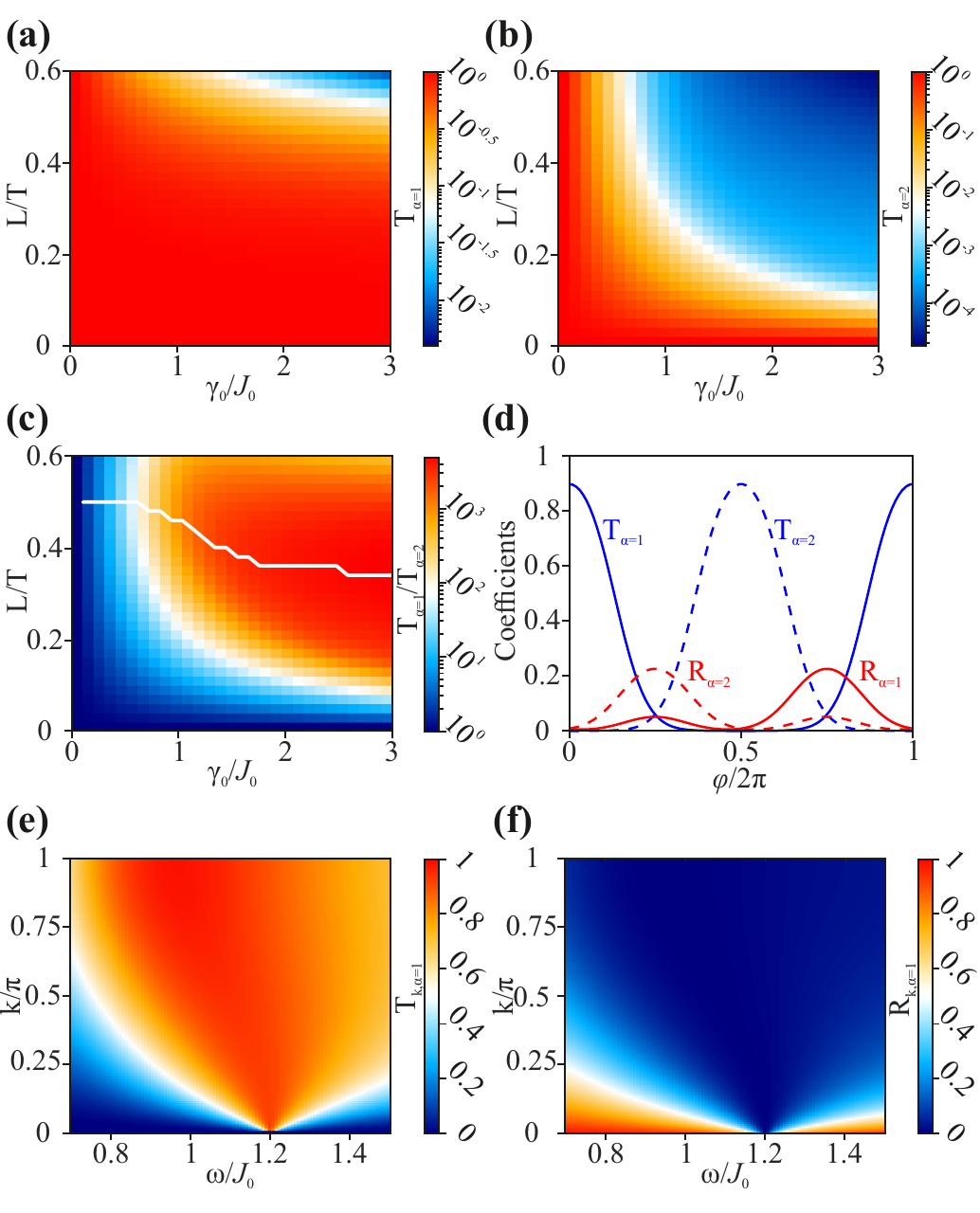}
\caption{Transmission of (a) right and (b) left movers over the impurity parameters $\gamma_0$ and $L$ at $\omega_0=1.195J_0$. (c) Decadic logarithm of $T_{\alpha=1}/T_{\alpha=2}$ for $\omega_0=1.195J_0$,
white line marks the function $\mathrm{max}_L(T_{\alpha=1}/T_{\alpha=2})(\gamma_0)$. (d) Transmission (blue) and Reflection (red)
of right (solid) of left (dashed) moving states over the shift angle $\varphi$ for $\gamma=1.5J_0$ and $L=0.25~T$. (e) Transmission and (f) reflection (f) of right movers for $L=0.25~T$
and $\gamma_0=1.5J_0$ over driving frequency $\omega$ and quasi-momentum $k$. }
 \label{fig:TheoryResults}
\end{figure}

We calculate transmission and reflection coefficients for scattering by the impurity with the use of Floquet-S-matrix theory (see Appendix~\ref{appendix C}). As a natural initial condition for the ratchet we assume a uniform superposition of either all the right- or left-moving states and consider band averaged quantities (Appendix \ref{appendix C}, eq.~\ref{eq:DissAveragedTandR}). Figs.~\ref{fig:TheoryResults} (a) and (b) display the transmission over the loss duration  $L$ and strength $\gamma_0$ at $\varphi=0$ for the $\alpha=1$ and $\alpha=2$ states, respectively.  We clearly see that for a wide range of system parameters our proposed scheme works as an excellent direction-dependent filter. This is prominently visible for $L \leq 0.5~T$ where the right-moving states are transmitted with $T_{\alpha=1}=\mathcal{O}(1)$, while the transmission of the left movers drops sharply with increasing $\gamma_0$ and $L$. Similar results are found in the disconnected dimer model in Appendix \ref{appendix B}, Eq.~(\ref{eq:DDDampedtrans}). The ratio $T_{\alpha=1}/T_{\alpha=2}$ [Fig.~\ref{fig:TheoryResults}~(c)] quantifies this effect and can exceed $10^{3}$ in the examined parameter range. Our illustrative picture of generating directional losses can be formalized by looking at the matrix elements of the impurity operator in the basis of the Floquet-Bloch states $V_{(q,\beta,n),(k,\alpha,m)}=1/T \int_0^T dt~\bra{\phi_{q,\beta,n}(t)} V(t) \ket{\phi_{k,\alpha,m}(t)}$, where the states $\ket{\phi_{k,\alpha,m}(t)}$ are calculated by (\ref{eq:DissEqninFloquetSpace}). The parts diagonal in the band index $\alpha,\beta=$\textcolor{blue}{$1,2$} determine the size of the transmission, while the off-diagonal ones couple the channels and induce reflection. Evaluating these integrals in our situation shows, that the matrix elements for the right movers  $V_{(q,1,n),(k,1,m)}$ and the inter-band coupling $V_{(q,1,n),(k,2,m)}$ are much smaller than for the left movers $V_{(q,2,n),(k,2,m)}$. This leads to the large suppression of the transmission of the left moving states while the right movers are almost unaffected. Using this picture explains the naively unexpected effect that the ratio $T_{\alpha=1}/T_{\alpha=2}$ has for all $\gamma_0$ a maximum at finite $L$: For small $L$ the dissipation of both directions is small leading to $T_{\alpha=1}/T_{\alpha=2}=\mathcal{O}(1)$. At large $L$ however, especially if $L>0.5~T$, also the right movers are strongly damped which implies a decrease of $T_{\alpha=1}/T_{\alpha=2}$. Here the time interval with losses is so long that they do not fit with sublattice oscillation of the right movers, leading to larger matrix elements and stronger damping. \\\\

We are interested in what happens if the parameters are tuned away from our proposed driving scheme and begin with changing the parameter $\varphi$. The resulting transmission and reflection coefficents are shown in Fig \ref{fig:TheoryResults} (d).
At the points $\varphi=0,\pi$ unidirectional transport is most favorable, as the dissipation peaks such that it only damps either left or right movers strongly. For intermediate values of $\varphi$, the losses are present during the intermediate part of the motion of the states [see Fig \ref{fig:Model} (d)], where both sub-lattices host population of substantial weight. This results in a lower ratio $T_{\alpha=1}/T_{\alpha=2}$ compared to $\varphi=0,\pi$ and confirms that we indeed took the optimal values for our proposed driving scheme.
The reflection coefficients have a maximum at $\varphi=\pi/2,3\pi/2$. Here, the dissipation is centered at the time, when both sub-lattices are populated equally, creating the largest matrix elements $V_{(q,\alpha,n),(k,\beta\not=\alpha,m)}$ and strongest coupling between the Remarkably, the reflection coefficients are not mirror symmetric about $\varphi=\pi$. We sketch this interesting feature for the case $\varphi=\pi/2$. Here the dissipation is timed such that a wave package consisting of right moving states is damped before it performs the sublattice oscillation inside the dissipative region, leading to small reflection. In case of a left moving package the dissipation is timed such that it delays the sublattice oscillation when the package is about to enter the lossy region. As a result, a major portion stays outside and thus gets reflected. For $\gamma_0 \to \infty$ and $L\approx 0.5~T$ this effect increases up to $R_{\alpha=2}\approx1$ and $R_{\alpha=1}\approx0$.

In Fig.~\ref{fig:TheoryResults} (e), (f) the transmission and reflection coefficients are plotted in dependence of quasi momenta $k$ and frequency $\omega$. As the coefficients are symmetric in $k$, we restrict ourselves to half the Brillouin zone, i.e.~$k>0$. 
The transmission is $\mathcal{O}(1)$ and homogeneous for all quasimomenta at the point of helical bands $\omega=\omega_{0}=1.1948~J_0$. Moving away from the ideal case, transmission around $k=0$ decreases due to hybridisation of the quasienergy bands  [see Fig.~1(b)] which mixes right and left moving states. These hybridized states do not perform a full oscillation between the sublattices, so both right and left movers are affected by the dissipation. This leads to a reduced transmission coefficient in comparison to the ideal case of helical bands. The reflection coefficient shown in Fig.~\ref{fig:TheoryResults} (f) is small and homogeneous for $\omega=\omega_{n=0}$, while it increases around $k=0$ due to the fact that the hybridized states lead to a finite inter-band coupling $V_{(q,\alpha,n),(k,\beta\not=\alpha,m)}$. In the motion of a wave package the states near $k=0$ are slow compared to others as the group velocity tends to zero at $k=0$. As these parts of the package reach the impurity at a late time in an experimental setting, our scheme filters out the slow parts in the transmitted package.

\section{Experiments}\label{sec:Experiments}

We realize unidirectional transmittance in arrays of coupled dielectric-loaded surface plasmon polariton waveguides (DLSPPWs). 
Here, we rely on the mathematical analogy between the  tight-binding Schrödinger equation and the paraxial Helmholtz equation which describes propagation of light in coupled waveguides~\cite{garanovich2012light,bleckmann2017spectral,cherpakova2017transverse}. According to this analogy, time is directly mapped into propagation distance which enables to mimic a Floquet system by periodic modulation of the corresponding parameters along the waveguide axis~\cite{rechtsman2013photonic,fedorova2019limits}. Precise control of the system's parameters including losses  as well as powerful detection technique make DLSPPWs an ideal system to investigate transmission through a region with local dynamic dissipation.

\begin{figure}[btbtbtb]
\includegraphics[width=\linewidth]{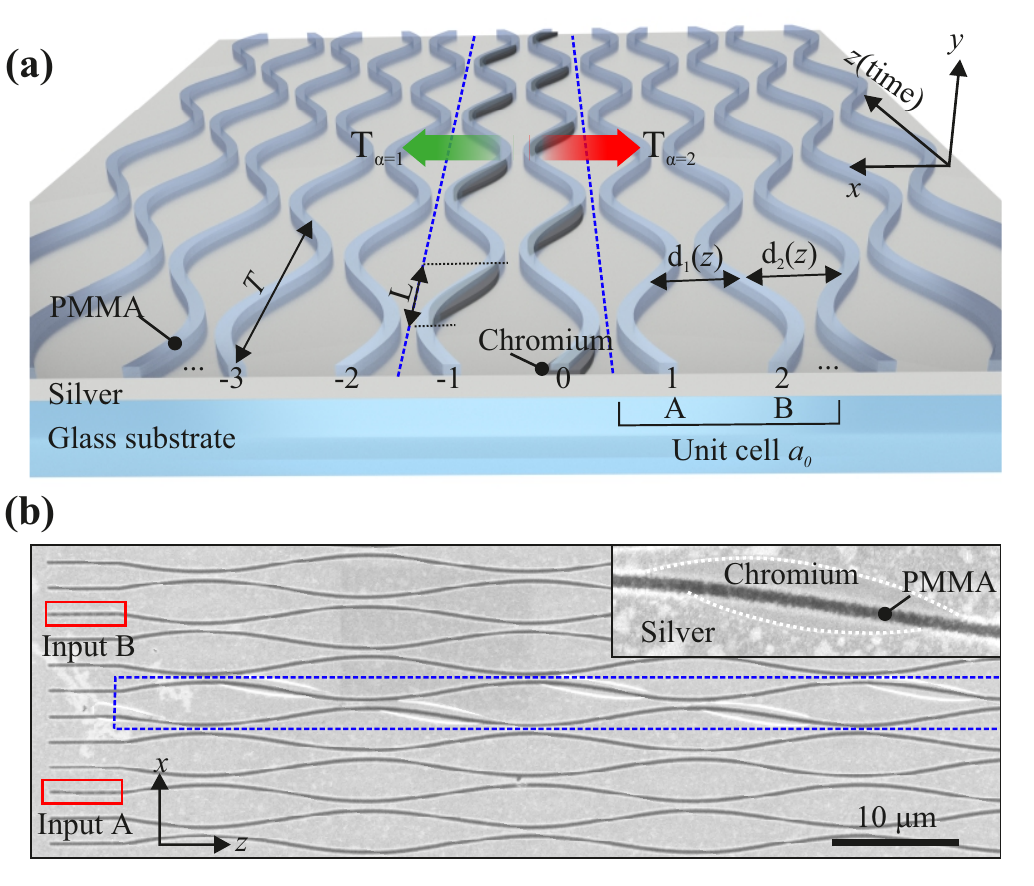}
\centering
\caption{(a) Sketch of a plasmonic  waveguide array featuring unidirectional transmittance. Green and red arrows indicate the low and high loss directions, respectively. (b) SEM scan of a typical sample. Inputs A and B are marked by red boxes and blue dotted line highlight the region with periodic dissipation. The chromium stripe used to implement losses is magnified in the top right corner.}
 \label{fig:Sample}
\end{figure}

A sketch as well as a scanning electron micrograph of a typical sample are shown in Figs.~\ref{fig:Sample}~(a) and (b), respectively. 
 See Appendix~\ref{appendix D} for sample fabrication and geometrical parameters of DLSPPWs. The array displayed in Fig.~\ref{fig:Sample}~(a) is analogous to a one-dimensional Floquet chain with two sites (waveguides) per unit cell, A and B. The sinusoidal modulation of the center-to-center distances $d_{1,2}(z)$  results in periodic modulation of the corresponding coupling constants. This modulation can be expressed by Eq.~\eqref{eq:Hopping} since the mode overlap decays as $\propto e^{-a\cdot d}$ with the distance $d$ between the waveguides. The parameters from Eq.~\eqref{eq:Hopping} were determined in an auxiliary experiment with just two waveguides by measuring the coupling length $L_{\mathrm{couple}}=\frac{2\pi}{C}$ in dependence on the distance between two waveguides $d$. Fitting the function $\ln C(d)$ by a line we obtain $\lambda=2.11\pm 0.21$ and $J_0=0.16\pm 0.05~\mathrm{\upmu m}^{-1}$. Due to strong confinement of SPPs we can neglect the variation of a propagation constant caused by waveguide bending and consider the real part of the on-site potential to be zero~\cite{fedorova2019limits}. 
We introduce local periodic dissipation by deposition of chromium stripes below the waveguides. Cr can cause strong losses with negligible effect on the real part of the effective refractive index~\cite{guo2009observation,song2019breakup}. Using simulations based on
finite-element analysis (COMSOL Multiphysics) we estimate the minimum loss strength induced by the Cr layer to be $\gamma_0=0.25~\mathrm{\upmu m}^{-1}$. The width of the Cr stripe is designed to be much larger than the width of a waveguide (see Fig.~\ref{fig:Sample}~(b)). We can, therefore, assume the losses to be approximately constant along the whole length of the stripe $L$ as given by Eq.~\eqref{eq:LossFunction}. 
We note that in addition to the engineered losses, the propagation of SPPs is accompanied by the constant decay rate $\beta''=(7.3\pm0.02)\cdot10^{-3}~\mathrm{\upmu m}^{-1}\ll\gamma_0$ caused by ohmic losses in the metal, imperfections of the fabricated film, and leakage radiation into the substrate. These losses are assumed to be homogeneous and independent of $z$. The propagation of SPPs in the array is monitored by real- and Fourier space leakage radiation microscopy (see Appendix~\ref{appendix E}).
\begin{figure}[bbbb]
\centering
\includegraphics[width=\linewidth]{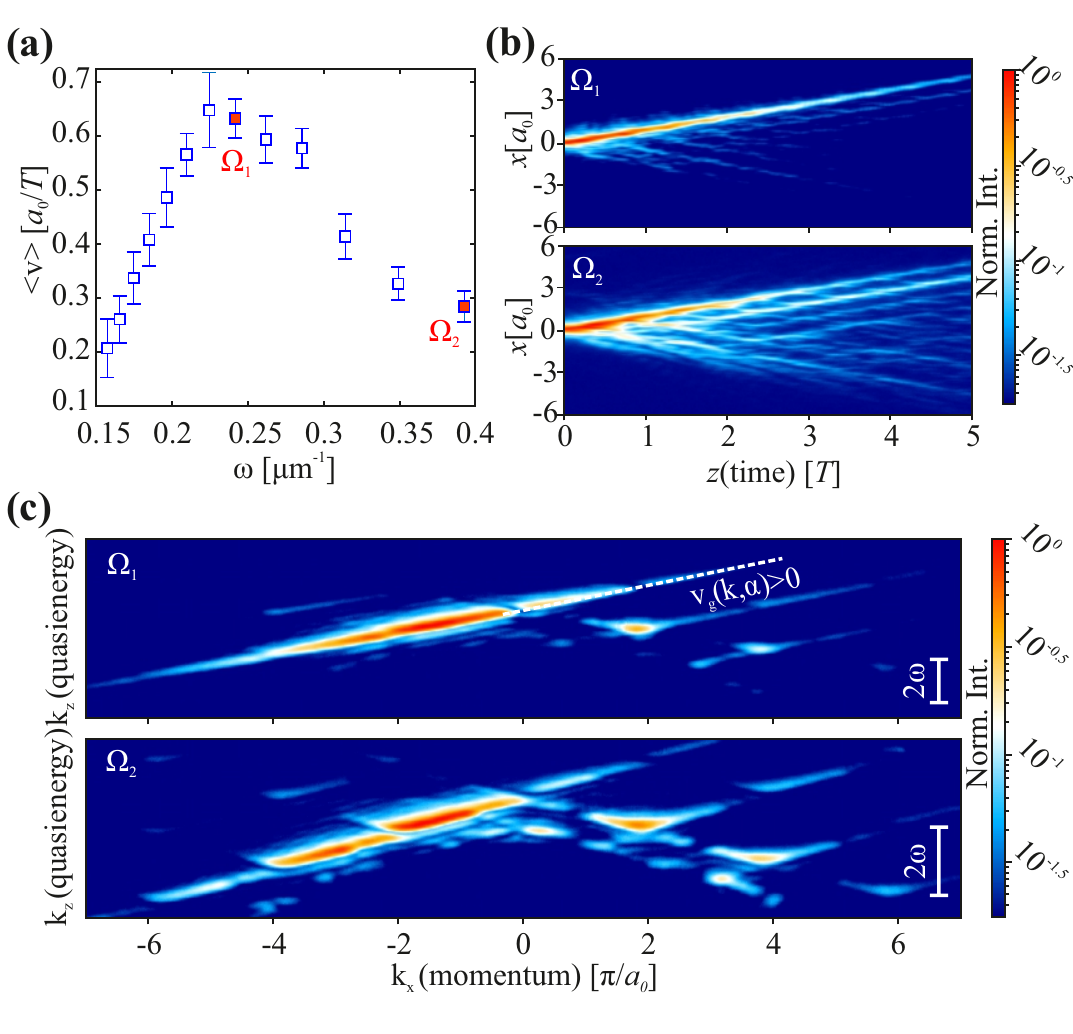}
\caption{(a) Measured mean group velocity of a wavepacket versus driving frequency $\omega$ for the single-site excitation at the input A. (b) Real-space SPP intensity distributions corresponding to the arrays modulated with frequency $\Omega_1=0.23~\mathrm{\upmu m}^{-1}$ and $\Omega_2=0.39~\mathrm{\upmu m}^{-1}$. (c) Fourier-space SPP intensity distributions for the same arrays as in (b). }
 \label{fig:Experimental1}
\end{figure} 

\begin{figure}[t]
\centering
\includegraphics[width=\linewidth]{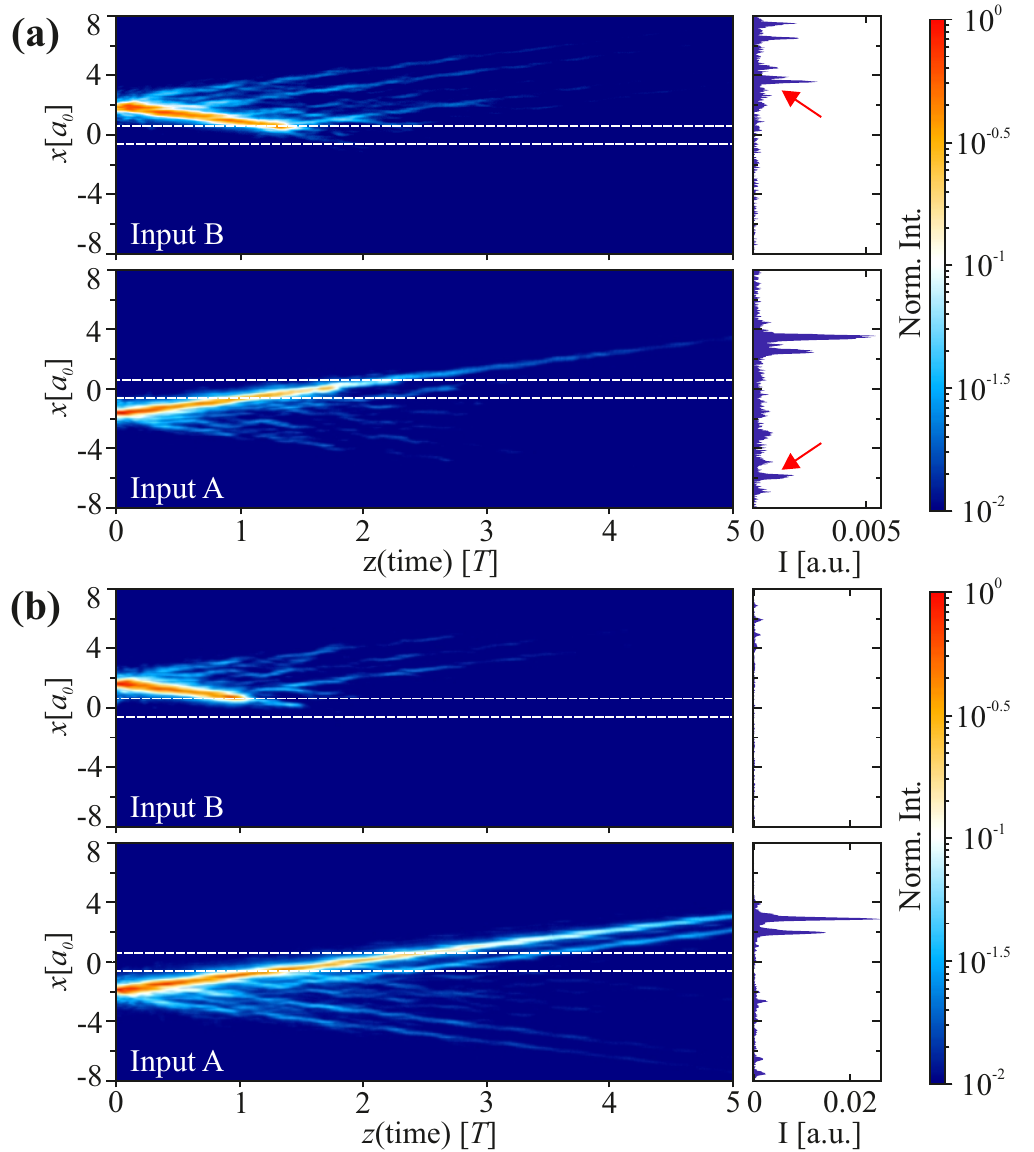}
\caption{Real space intensity distributions for the DLSPPW arrays with local modulated dissipation (highlighted by white dashed lines) featuring unidirectional transmittance at $\Omega_1$. The wavepacket is excited at $x>0$ (input B, top) or at $x<0$ (input A, bottom). The area plots at the right side from the real-space data show the intensity distribution after the propagation distance $z=5T$. (a) Chromium stripes with $L=0.3T$ were deposited below two waveguides. The red arrows point to reflected wave. (b) The same as in (a), but the length of the Cr stripe was reduced to $L=0.15T$.}
 \label{fig:Experimental2}
\end{figure} 

First, we consider the case without engineered losses and determine the driving frequency $\omega=2\pi/T$ at which
we can achieve directed transverse motion of SSPs with the highest group velocity $v_{\mathrm{g}}$. Theory predicts for this case the absence of hybridization of the counter-propagating states and therefore the most pronounced one-way transmission effect. To find the group velocity maximum we fabricate arrays of modulated DLSPPWs (no Cr is deposited) with various frequencies of modulation. For every array we measure the real-space intensity distribution $I(x,z)$ (analogous to $|\Psi (x,t)|^{2}$) after the single-site excitation at the sublattice A. Note that the corresponding data for the input B is just mirrored about $x=0$. We use the experimental data to extract the position of the center of mass (CoM) of the wavepacket $\langle x\rangle (z)=\sum_j I(x_j,z)\cdot x_j/\sum_j I(x_j,z)$ as a function of $z$. The group velocity $v_{\mathrm{g}}$ is found as the slope of the linear fit to $\langle x\rangle (z)$ and plotted in units of a unit cell per driving period $a_0/T$ against $\omega$ in Fig.~\ref{fig:Experimental1}~(a).  The resulting curve reaches the peak value of about $0.63$ at $\omega=\Omega_{1}\approx 0.23~\mathrm{\upmu m}^{-1}$. We note that the measured peak value of the group velocity is smaller than $1$ as would be expected from the completely filled band (see Fig.~\ref{fig:Model}~(c)). We attribute this deviation to the contribution of camera noise and non-perfect excitation conditions. By the latter we mean that, first, the overlap of the states moving in the $-x$ direction with the initial conditions is not exactly zero, second, when guided SPPs are excited by shining laser light onto the grating coupler, the laser spot slightly excites the neighboring waveguides. These factors inevitably decrease the CoM displacement.
The quantitative comparison with the theoretical value of $\omega_0$ in Eq.~(\ref{eq:DissConditionLinearBands}) requires the value of $J_0$, which has an experimental uncertainty $1.09 J_0 \alt \Omega_1 \alt 2.09 J_0$ that is fully consistent with the theoretical value of $\omega_0 = 1.195 J_0$.  In Figs.~\ref{fig:Experimental1}~(b), (c)  we compare the real- and Fourier-space intensity distributions for two frequencies $\Omega_1$ (close to the resonance $\omega_0$) and $\Omega_2$ (maximum frequency in our measurements, away from the resonance). In real space at $\Omega_1$ we observe that the wavepacket is confined,  and the intensity maximum is transported in  positive $x$-direction [Fig.~\ref{fig:Experimental1}~(b), top]. The corresponding Fourier intensity distribution $I(k_x,k_z)$ shown in Fig.~\ref{fig:Experimental1}~(c) (top) reveals nearly linear dispersion and predominant population of the Floquet states with $v^{\mathrm{g}}_{k,\alpha}>0$. In contrast, at $\Omega_2$ the wavepacket is spreading in both directions [see Fig.~\ref{fig:Experimental1}~(b), bottom] and in Fourier space [see Fig.~\ref{fig:Experimental1}~(c), bottom] the gaps between the quasienergy bands broaden, and the states with $v^{\mathrm{g}}_{k,\alpha}<0$ become noticeably populated.
Such a behaviour results from hybridization of counter-propagating states and is fully consistent with the theory [compare with Fig.~\ref{fig:Model}~(b)].

Next, we fabricate DLSPPW arrays with local modulated losses using the optimal driving frequency $\Omega_{\mathrm{1}}$ determined above. For that we deposit Cr stripes of length $L=0.3T$ beneath the two waveguides in between the inputs A and B, so that the  phase shift in Eq.~(\ref{eq:LossFunction}) is zero $\varphi=0$ (see Fig.~\ref{fig:Model}~(b)). The inputs are placed such that the excited wavepacket impinges upon the region of modulated losses from both sides. In Fig.~\ref{fig:Experimental2}~(a) the resulting real-space intensity distributions of SPPs for two input conditions are displayed. Here, the wavepacket impinging from the region $x>0$ (top image) is strongly damped such that no SPPs are visible after the lossy region. Since the transmitted wave is lower than the noise level, the transmission coefficient must be $T_{\alpha=2}<10^{-2}$. In contrast, when the wavepacket impinges from the opposite side, $x<0$, it is partially transmitted (bottom image). By comparing to the case with no loss, we can estimate the transmission coefficient $T_{\alpha=1}\approx0.53$. Additionally, the weak reflection from the interface is observed for both sides (see red arrows). This can be related to the slight shift of the Cr stripes in respect to the waveguides which leads to a non-zero phase shift $\varphi$.

We now aim to improve the performance of the direction-dependent filtering in our system, in particular, we want to increase the transmission in the low-loss direction $T_{\alpha=1}$ while keeping the $T_{\alpha=2}$ below the detection limit of $\sim 10^{-2}$. Relying on the numerical calculations discussed above, at the constant loss strength $\gamma_0$ this can be realized by reducing the Cr stripe length $L$. Indeed, for $L=0.15T$ (Fig~\ref{fig:Experimental2}~(b)) we again observe strong absorption in $-x$ direction such that  $T_{\alpha=2}<10^{-2}$, however, the transmission in the opposite direction is substantially increased $T_{\alpha=1}\approx0.92$. In this case we see no reflection from the interface.

\section{Conclusion}\label{sec:Conclusion}

In conclusion, we proposed and experimentally realized a novel, direction-dependent filter in a Hamiltonian quantum ratchet. Our ratchet scheme  on a 1D periodically driven lattice is inspired by the arrays of evanescently coupled waveguides as experimental platform. Using Floquet-Bloch theory we show that at certain frequencies such a model supports directional transport characterized by helical Floquet bands with negligible dispersion. Based on the sublattice oscillation of the Floquet-Bloch states we introduce local periodic losses as a new method for direction-dependent filtering. In doing so we achieve strong non-reciprocal transport at all quasi-momenta and a vast range of system parameters. 
In order to quantify non-reciprocal transmission, we develop the Floquet scattering theory for the conceptually interesting though commonly undiscussed case where both bulk and scattering potentials are modulated time-periodically. Using this approach, we calculate asymmetric transmission and reflection coefficients for various system parameters and determine the optimal conditions for direction-dependent filtering. Based on the theoretical predictions we realize our ratchet model in arrays of coupled periodically modulated plasmonic waveguides. Using real- and Fourier space measurements we determine the resonant modulation frequency corresponding to the highest group velocity and almost dispersionless bands. Next, the non-Hermitian impurity is implemented by means of an absorber deposited locally below the waveguides in a two-step lithographic process. Non-reciprocal transmission through this impurity is clearly demonstrated by real-space intensity distributions. 
Our results indicate that it is possible to create a Hamiltonian ratchet being intrinsically non-reciprocal such that any mixed initial state can be filtered to achieve motion in only one chosen direction. This exceeds the known techniques in a novel way. Contrary to non-reciprocal transport induced by non-Hermitian gauge fields no additional gain is needed~\cite{Longhi2015}. This makes our method favourable in further experimental settings such as ultracold quantum gases~\cite{ni2017hamiltonian,denisov2007periodically}.

\appendix
\section{Floquet-Bloch theory for the Hamiltonian ratchet}\label{appendix A}

We study our ratchet model within a framework of the Floquet-Bloch theory~\cite{gomez2013floquet}. For that we first transform the bulk Hamiltonian from Eq.~(\ref{eq:BulkHamiltonian}) to $k$-space
	\begin{equation}
	H_{\rm Bulk}(t)=\sum_k \psi_k^\dagger H_k(t) \psi_k.
	\end{equation}
Here, we introduced $\psi_k=\frac{1}{\sqrt{L}} \sum_j e^{-\mathrm{i} k j} \psi_j$ where $\psi_j=(c_{j,\rm A},c_{j,\rm B})^T$ and $H_k(t)$ is the time-dependent Bloch Hamiltonian which for our model reads
\begin{equation}
	\label{eq:DissHailtonainkspase}
	H_k(t)=\begin{pmatrix}
	0 & h(k) \\ h^*(k) & 0
	\end{pmatrix},
\end{equation}
where $h(k)=J_1(t)e^{\mathrm{i} k a_0/2}+J_2(t)e^{-\mathrm{i} k a_0/2}$ and $^*$ denotes the complex conjugation.
The steady states of the Hamiltonian (\ref{eq:DissHailtonainkspase}) are the so called Floquet-Bloch states~\cite{gomez2013floquet}
\begin{equation}
    \label{eq:DissFloquetBlochStates}
    \ket{\psi_{k,\alpha}(t)}=e^{-\mathrm{i} \epsilon_{k,\alpha,m} t} \ket{\phi_{k,\alpha,m}(t)},
\end{equation}
which are comprised of a phase factor involving the quasienergy $\epsilon_{k,\alpha,m}$ and the $T$-periodic Floquet mode $\ket{\phi_{k,\alpha,m}(t)}$. The index $m$ arises from periodicity of the quasienergies, namely $\epsilon_{k,\alpha,m}=\epsilon_{k,\alpha,0}+m \omega$, where $ \epsilon_{k,\alpha,0}$ lies in the so-called first Floquet-Brillouin zone $[-\omega/2,\omega/2)$. As the corresponding modes only differ by a phase factor, $\ket{\phi_{k,\alpha,m}(t)}=e^{i m \omega t}\ket{\phi_{k,\alpha,0}(t)}$, the solutions from different Floquet-Brillouin zones describe the same physics. In our case the states can be labeled by their quasi-momentum $k$, the band  $\alpha=1,2$, and  the Floquet index $m$. 
The Floquet modes are eigenstates of the Floquet operator $\mathcal{H}_k(t)=H_k(t)-i \frac{\partial}{\partial t}$ in Floquet space $\mathcal{F}=\mathcal{R}\otimes \mathcal{T}$, which consists of the configuration space $\mathcal{R}$ and the space of the time-periodic functions $\mathcal{T}$ \cite{Eckardt2017,Sambe1973}. This eigenvalue equation reads
\begin{equation}
    \label{eq:DissEqninFloquetSpace}
    \hat{\mathcal{H}}_k\fket{\phi_{k,\alpha,m}}=\epsilon_{k,\alpha,m} \fket{\phi_{k,\alpha,m}}.
\end{equation}
Here $\fket{\phi}$ is the element of $\mathcal{F}$ which corresponds to $\ket{\phi(t)}$ and $\hat{A}$ is the operator which is acting in $\mathcal{F}$ connected with $A(t)$. Introducing the Fourier basis $\tket{n} \in \mathcal{T}$ and expressing $\tket{n(t)}=e^{-\mathrm{i} n \omega t}$, where $\ket{\phi_{k,\alpha,m}(t)}$ and $H_k(t)$ are written by their Fourier coefficients $f^{(n)}=1/T \int_0^T dt~e^{\mathrm{i} n \omega t} f(t)$,  Eq.~(\ref{eq:DissEqninFloquetSpace}) transforms to
\begin{equation}
\label{eq:DissFloquetHamultonian}
    \sum_m[H_k^{(n-m)}-n \omega \delta_{n,m}] \ket{\phi_{k,\alpha,m}^{(m)}}=\epsilon_{k,\alpha,m}\ket{\phi_{k,\alpha,n}^{(m)}}.
\end{equation}
Eq.~(\ref{eq:DissFloquetHamultonian}) is first truncated in Floquet space, the resulting eigenvalue problem involving a finite matrix is solved numerically in an efficient manner.

\section{Disconnected Dimer Model}\label{appendix B}
 In order to qualitatively understand the origin of the rectified transport,  we consider the Hamiltonian (2)  with a simplified driving scheme, where one period, for $0\leq t<T$, reads:
 \begin{subequations}
  \label{eq:DissSimpleCoupling}
     \begin{align}
            J_1(t) =
  \begin{cases}
    J, & \quad t\in [t_1,t_1+\delta t[ \\
    0,  & \quad \text{otherwise}
  \end{cases},
\\
  J_2(t) =
  \begin{cases}
    J, & \quad t\in [t_2,t_2+\delta t[\\
    0,  & \quad \text{otherwise}
  \end{cases}. 
     \end{align}
 \end{subequations}

Here $t_1+\delta t<T/2$ with $\delta t > 0$ and $t_2=T/2+t_1$. The Schrödinger equation for $\ket{\psi (t)}=\sum_j( \psi_j^{\rm A} c_{j,\rm A}^\dagger+\psi_j^{\rm B} c_{j,\rm B}^\dagger)\ket{0}$ reads
\begin{subequations}
	\begin{eqnarray}
	 	i \partial_t \psi_j^{\rm A}&=&J_1(t) \psi_j^{\rm B}+J_2(t) \psi_{j-1}^{\rm B},\\
	 	i \partial_t \psi_j^{\rm B}&=&J_1(t) \psi_j^{\rm A}+J_2(t) \psi_{j+1}^{\rm A}, 
	\end{eqnarray}
\end{subequations}
where we assume as initial conditions $\psi_j^{\gamma}(t=0)=\delta_{\gamma,\rm A}\delta_{j,l}$. For $t \in [0,t_1[$ all couplings are zero and the wave function stays at its initial value. In the interval $t \in [t_1,t_1+\delta t[$ the state undergoes Rabi oscillations between sublattice A and B
 \begin{subequations}
 	\begin{align}
    &\psi_l^{\rm A}(t)=\cos[J (t-t_1)],\\
 	&\psi_l^{\rm B}(t)=-i \sin[J (t-t_1)].
 	\end{align}
\end{subequations}
We assume in the following that population is fully transferred between the sublattices, i.e. the condition $J \delta t=\frac{\pi}{2}+n \pi$ holds with $n \in \mathbb{N}_0$. A similar analysis for the second half of the cycle (\ref{eq:DissSimpleCoupling}) yields that the state moves in one driving-period $T$ one unit cell to the right $\psi_j^\gamma(t=T)=\delta_{\gamma,\rm A}\delta_{j,l+1}$, while a state starting on sublattice B moves the same length to the left. Thus we created a simple scheme for ideal, rectified transport. The quasienergy bands are linear $\epsilon_{k,\alpha,m}=(-1)^{\alpha-1}(\frac{\omega k}{2 \pi}-\frac{\omega}{2})+m \omega$. Using $\delta t=\delta \xi ~2\pi/\omega$, we find the driving frequencies where perfect linear bands occur to be
\begin{equation}
    \label{eq:DissCondLinearBandsModel}
    \omega_n=\frac{4 J \delta \xi}{1+2 n},n \in \mathbb{N}_0.
\end{equation}
As Eq.~(\ref{eq:DissCondLinearBandsModel}) only dependes on the area $J \delta \xi$, it can be generalized to eq.~(\ref{eq:DissConditionLinearBands}). Also in this model unidirectional transport can be investigated when an inpurity is added. We therefore look at a special scheme for the dissipative impurity potential Eq.~(\ref{eq:ModelPotential})
\begin{subequations}
\label{eq:DD}
    \begin{align}
       &\gamma_{\rm A}(t) =
  \begin{cases}
    \gamma_0, & \quad t\in [0,t_1[ \cup [t_2+\delta t,T[\\
    0,  & \quad \text{otherwise}
  \end{cases},
\\
 & \gamma_{\rm B}(t) =
  \begin{cases}
    \gamma_0, & \quad t\in [t_1+\delta t,t_2[\\
    0,  & \quad \text{otherwise}
  \end{cases}.  
    \end{align}
\end{subequations}

The right-moving state can transmit unaffected through the impurity, while the left moving state is exponentially damped. The transmission coefficients read
   \begin{subequations}
       	\label{eq:DDDampedtrans}
 	\begin{align}
 	& T_{\alpha=1}=1\\
    &T_{\alpha=2}=\exp[-2\gamma_0 L_{\rm DD}],
 	\end{align}
\end{subequations}
where $L_{\rm DD}=T-2\delta t$.

\section{Floquet Scattering Theory}\label{appendix C}
We are interested in the scattering properties of the Floquet-Bloch states $\fket{\phi_{k,\alpha,m}}$ in the presence of an in general non-Hermitian impurity operator $V(t)$. Contrary to the common literature \cite{Millack1990,Moskalets2002,Smith2015,Li2018,Li2018a}, we look at a setup where both bulk and scattering potential are modulated time-periodically with the same frequency $\omega$. It turns out that the Lippmann-Schwinger equation for Floquet systems equals a static Lippmann-Schwinger theory in the Floquet space $\mathcal{F}$. 
 It is central to calculate the matrix elements of the Floquet S-matrix in the Floquet-Bloch basis $S_{(q,\beta,n),(k,\alpha,m)}$. For a general operator $A(t)$ these matrix elements are defined by $
A_{(q,\beta,n),(k,\alpha,m)}=\frac{1}{T} \int_0^T dt~\bra{\phi_{q,\beta,n}(t)} A(t) \ket{\phi_{k,\alpha,m}(t)}$, where the states $\ket{\phi_{k,\alpha,m}(t)}$ are calculated by (\ref{eq:DissEqninFloquetSpace}).  The matrix elemntes of the Floquet S-matrix describe the amplitude for finding a particle in the state $(q,\beta,n)$ after the scattering process if initially it was in state $(k,\alpha,m)$. The Floquet-S matrix can be termed as
    \begin{equation}\centering
    \label{eq:DissFloquetSmatrix}
    \begin{split}
        S_{(q,\beta,n),(k,\alpha,m)}=\delta(k-q)\delta_{\alpha,\beta}\delta_{n,m}-\\-2 \pi i \delta(\epsilon_\beta-\epsilon_\alpha) T_{(q,\beta,n),(k,\alpha,m)},
    \end{split}
	\end{equation}
	where we introduce the matrix elements of the Floquet-T matrix $\hat{T}=\hat{V}\hat{\Omega}^+$, with the impurity operator $\hat{V} \in \mathcal{F}$ and the M{\o}ller operator $\hat{\Omega}^+=\mathbb{I}+(\epsilon+i0^+-\hat{\mathcal{H}}_{\rm bulk}-\hat{V})^{-1}\hat{V}$.
	The matrix elements of the Floquet-T matrix are given by the self-consistency equation
	\begin{equation}
	\label{eq:DissLSTmatrixeqnkbasis}
	\begin{split}
	T_{(q,\beta,n),(k,\alpha,m)}=V_{(q,\beta,n),(k,\alpha,m)}+\\+\sum_{\delta,l} \int_{-\pi}^{\pi} dp \frac{V_{(q,\beta,n),(p,\delta,l)}}{\epsilon_{k,\alpha,m}-\epsilon_{p,\delta,l}+i0^+} 
	T_{(p,\delta,l),(k,\alpha,m)}.
	\end{split}
	\end{equation}
	 We transform Eq.~(\ref{eq:DissLSTmatrixeqnkbasis}) to a linear system which is solved numerically by discretizing the quasi-momentum space and by introducing a cuttoff $m_{\rm max}$ in the Floquet index.
In our case, where bulk and scattering potential are driven by the same frequency, Eq.~(\ref{eq:DissFloquetSmatrix}) dictates that the scattered waves reside in the same Floquet-Brillouin zone as the incoming wave, while all other channels host evanescent waves. Thus only a $2\times2$ sub-matrix
\begin{equation}
 \tilde{S}_{\rm scatt}(k)=\begin{pmatrix} t_{k,\alpha=1} & r_{-k,\alpha=2} \\ r_{k,\alpha=1} & t_{-k,\alpha=2}. \end{pmatrix}   
\end{equation} of the Floquet-S Matrix will be non-zero and contribute to scattering. Here we introduced the $k$-dependent transmission amplitude $t_{k,\alpha}=1-\frac{2 \pi i}{|v_{k,\alpha}^{\rm g}|}T_{(k,\alpha,m),(k,\alpha,m)}$ and reflection amplitude $r_{k,\alpha}=- \frac{2 \pi i}{|v_{k,\alpha}^{\rm g}|}T_{(-k,\beta\not=\alpha,m),(k,\alpha,m)}$ by the non-singular parts of the Floquet-S matrix. The $k$-dependent transmission and reflection coefficients measuring the probability for these events are given by $T_{k,\alpha}=|t_{k,\alpha}|^2$ and $R_{k,\alpha}=|r_{k,\alpha}|^2$.
  In our setting we average these quantities over a full band
   \begin{subequations}
   \label{eq:DissAveragedTandR}
	\begin{eqnarray}
       T_\alpha=\frac{1}{2 \pi} \int_0^{2 \pi}dk T_{k,\alpha}, \\
       R_\alpha=\frac{1}{2 \pi} \int_0^{2 \pi}dk R_{k,\alpha}.
 	\end{eqnarray}
\end{subequations} In the case of a Hermitian impurity, the matrix $\tilde{S}_{\rm scatt}(k)$ is an unitary matrix \cite{Moskalets2002}, implying the relation $T_{k,\alpha=1}=T_{-k,\alpha=2}$. This results in equal averages $T_{\alpha=1}=T_{\alpha=2}$ and shows that non-reciprocal transport is not possible in the hermitian case.

\section{Sample fabrication}\label{appendix D}
The DLSPPW arrays with locally modulated dissipation are fabricated with a two-step electron beam lithography process (EBL). 
The sample preparation starts with evaporation of 62~nm of Ag and 2~nm of Cu for adhesion on a cleaned surface of a glass substrate. Then the sample is spin-coated with the polymeric resist poly(methyl methacrylate) (PMMA). In the first EBL step, we utilize PMMA as a positive-tone resist in order to fabricate a template for the lossy regions and alignment markers. The areas exposed to the electron beam are dissolved in a developer and 15 nm of Cr is evaporated on top of the substrate. After the lift-off process we end up with the Cr stripes and the alignment markers at the predefined positions. The width of each Cr stripe is set to $1.3~\mathrm{\upmu m}$.  Then the sample is again spin-coated with PMMA and the second EBL step takes place. Now we fabricate the DLSPPW arrays on top of the Cr stripes using the markers for the alignment. In this step PMMA acts as a negative tone resist which is achieved by increasing the applied electron dose~\cite{block2014bloch}. Finally, the samples are developed in acetone.
The atomic force microscopy measurements revealed that the applied electron dose results in the mean waveguide height of $90$~nm and the width of $270$~nm which allows us to work in a single-mode regime at a vacuum wavelength of $\lambda=0.98~\mathrm{\upmu m}$. For these geometrical parameters the propagation constant of the guided mode is $\beta=\beta'-i\beta''=\mathrm{const}$, $\beta'=6.55~\mathrm{\upmu m}^{-1}$ (obtained by numerical simulations with COMSOL Multiphysics) and $\beta''=(7.3\pm0.02)\cdot10^{-3}~\mathrm{\upmu m}^{-1}$ (obtained by measuring propagation length of SPPs).
The distance between the adjacent waveguides varies as $d_{1,2}=2 \pm 0.65\cdot\sin{\omega z} ~\mathrm{\upmu m}$, $\omega=2\pi/T$. As shown in Fig.~\ref{fig:Sample}~(b) the modulated part of the array is preceded by a short straight interval of the length $6~\mathrm{\upmu m}$. This region contains the grating coupler (red box) which is used for SPP excitation. The grating is deposited only onto the two waveguides at the left and right side from the dissipative region (inputs A and B),  while the extension of others to this region is needed to prevent fire-end excitation of the adjacent waveguides.

\section{Leakage Radiation Microscopy}\label{appendix E}
SPPs are excited by focusing a TM-polarized laser beam with $\lambda_0$=980~nm  (NA of the focusing objective is 0.4) onto the grating coupler deposited on top of the chosen waveguide. The propagation of SPPs in the array is monitored by real- and Fourier-space leakage radiation microscopy (LRM)~\cite{cherpakova2017transverse,fedorova2020observation}. The leakage radiation as well as the transmitted laser beam are both collected by a high NA oil immersion objective (Nikon 1.4 NA, 60x Plan-Apo). The transmitted laser was filtered out by placing a knife edge at the intermediate back focal plane (BFP) of the oil immersion objective. The remaining radiation was imaged onto an sCMOS camera (Andor Marana). Real-space SPP intensity distributions were recorded at the real image plane while the momentum-space intensity distribution was obtained by imaging the BFP of the oil immersion objective.

\section*{AUTHOR CONTRIBUTIONS}

ZF and CD contributed equally to this work. ZF fabricated the samples, conducted the experiments, contributed to the theoretical understanding of the effects, and prepared the figures. CD analysed the system theoretically, developed the numerical method based on the Floquet-S-Matrix theory and performed calculations of the transmission and reflection coefficients. ZF and CD wrote the first draft of the manuscript. AS contributed to the theoretical understanding of the ratchet effect and calculated the Floquet spectrum. SL conceived the plasmonic experiment and supervised ZF and AS. SE and JK conceived the theoretical analysis. SE  supervised CD and ZF,  JK supervised ZF. SL, SE, and JK contributed to the manuscript writing. All authors discussed the results and reviewed the manuscript.

\begin{acknowledgments}
We acknowledge financial support by the Deutsche Forschungsgemeinschaft through 
CRC/TR 185 (277625399) OSCAR (C.D., S.E., Z.F, J.K., S.L., A.S.) and through the Cluster of Excellence ML4Q (90534769), (Z.F., J.K., S.L., A.S.).
\end{acknowledgments}

\bibliographystyle{apsrev4-1}
\bibliography{bibliography}

\begin{thebibliography}{44}%
\makeatletter
\providecommand \@ifxundefined [1]{%
 \@ifx{#1\undefined}
}%
\providecommand \@ifnum [1]{%
 \ifnum #1\expandafter \@firstoftwo
 \else \expandafter \@secondoftwo
 \fi
}%
\providecommand \@ifx [1]{%
 \ifx #1\expandafter \@firstoftwo
 \else \expandafter \@secondoftwo
 \fi
}%
\providecommand \natexlab [1]{#1}%
\providecommand \enquote  [1]{``#1''}%
\providecommand \bibnamefont  [1]{#1}%
\providecommand \bibfnamefont [1]{#1}%
\providecommand \citenamefont [1]{#1}%
\providecommand \href@noop [0]{\@secondoftwo}%
\providecommand \href [0]{\begingroup \@sanitize@url \@href}%
\providecommand \@href[1]{\@@startlink{#1}\@@href}%
\providecommand \@@href[1]{\endgroup#1\@@endlink}%
\providecommand \@sanitize@url [0]{\catcode `\\12\catcode `\$12\catcode
  `\&12\catcode `\#12\catcode `\^12\catcode `\_12\catcode `\%12\relax}%
\providecommand \@@startlink[1]{}%
\providecommand \@@endlink[0]{}%
\providecommand \url  [0]{\begingroup\@sanitize@url \@url }%
\providecommand \@url [1]{\endgroup\@href {#1}{\urlprefix }}%
\providecommand \urlprefix  [0]{URL }%
\providecommand \Eprint [0]{\href }%
\providecommand \doibase [0]{http://dx.doi.org/}%
\providecommand \selectlanguage [0]{\@gobble}%
\providecommand \bibinfo  [0]{\@secondoftwo}%
\providecommand \bibfield  [0]{\@secondoftwo}%
\providecommand \translation [1]{[#1]}%
\providecommand \BibitemOpen [0]{}%
\providecommand \bibitemStop [0]{}%
\providecommand \bibitemNoStop [0]{.\EOS\space}%
\providecommand \EOS [0]{\spacefactor3000\relax}%
\providecommand \BibitemShut  [1]{\csname bibitem#1\endcsname}%
\let\auto@bib@innerbib\@empty
\bibitem [{\citenamefont {Denisov}\ \emph {et~al.}(2014)\citenamefont
  {Denisov}, \citenamefont {Flach},\ and\ \citenamefont
  {H{\"a}nggi}}]{denisov2014tunable}%
  \BibitemOpen
  \bibfield  {author} {\bibinfo {author} {\bibfnamefont {S.}~\bibnamefont
  {Denisov}}, \bibinfo {author} {\bibfnamefont {S.}~\bibnamefont {Flach}}, \
  and\ \bibinfo {author} {\bibfnamefont {P.}~\bibnamefont {H{\"a}nggi}},\
  }\bibfield  {title} {\enquote {\bibinfo {title} {Tunable transport with
  broken space--time symmetries},}\ }\href@noop {} {\bibfield  {journal}
  {\bibinfo  {journal} {Physics Reports}\ }\textbf {\bibinfo {volume} {538}},\
  \bibinfo {pages} {77} (\bibinfo {year} {2014})}\BibitemShut {NoStop}%
\bibitem [{\citenamefont {Smoluchowski}(1927)}]{smoluchowski1927experimentell}%
  \BibitemOpen
  \bibfield  {author} {\bibinfo {author} {\bibfnamefont {M.}~\bibnamefont
  {Smoluchowski}},\ }\bibfield  {title} {\enquote {\bibinfo {title}
  {Experimentell nachweisbare, der {\"u}blichen thermodynamik widersprechende
  molekularph{\"a}nomene},}\ }\href@noop {} {\bibfield  {journal} {\bibinfo
  {journal} {Pisma Mariana Smoluchowskiego}\ }\textbf {\bibinfo {volume} {2}},\
  \bibinfo {pages} {226} (\bibinfo {year} {1927})}\BibitemShut {NoStop}%
\bibitem [{\citenamefont {Feynman}\ \emph {et~al.}(1963)\citenamefont
  {Feynman}, \citenamefont {Leighton},\ and\ \citenamefont
  {Sands}}]{feynman1963feynman}%
  \BibitemOpen
  \bibfield  {author} {\bibinfo {author} {\bibfnamefont {R.}~\bibnamefont
  {Feynman}}, \bibinfo {author} {\bibfnamefont {R.}~\bibnamefont {Leighton}}, \
  and\ \bibinfo {author} {\bibfnamefont {M.}~\bibnamefont {Sands}},\ }\bibfield
   {title} {\enquote {\bibinfo {title} {The feynman lectures on physics, vol. 1
  addison wesley},}\ }\href@noop {} {\bibfield  {journal} {\bibinfo  {journal}
  {Reading, MA}\ } (\bibinfo {year} {1963})}\BibitemShut {NoStop}%
\bibitem [{\citenamefont {Mahmud}\ \emph {et~al.}(2009)\citenamefont {Mahmud},
  \citenamefont {Campbell}, \citenamefont {Bishop}, \citenamefont {Komarova},
  \citenamefont {Chaga}, \citenamefont {Soh}, \citenamefont {Huda},
  \citenamefont {Kandere-Grzybowska},\ and\ \citenamefont
  {Grzybowski}}]{mahmud2009directing}%
  \BibitemOpen
  \bibfield  {author} {\bibinfo {author} {\bibfnamefont {G.}~\bibnamefont
  {Mahmud}}, \bibinfo {author} {\bibfnamefont {C.~J.}\ \bibnamefont
  {Campbell}}, \bibinfo {author} {\bibfnamefont {K.~J.}\ \bibnamefont
  {Bishop}}, \bibinfo {author} {\bibfnamefont {Y.~A.}\ \bibnamefont
  {Komarova}}, \bibinfo {author} {\bibfnamefont {O.}~\bibnamefont {Chaga}},
  \bibinfo {author} {\bibfnamefont {S.}~\bibnamefont {Soh}}, \bibinfo {author}
  {\bibfnamefont {S.}~\bibnamefont {Huda}}, \bibinfo {author} {\bibfnamefont
  {K.}~\bibnamefont {Kandere-Grzybowska}}, \ and\ \bibinfo {author}
  {\bibfnamefont {B.~A.}\ \bibnamefont {Grzybowski}},\ }\bibfield  {title}
  {\enquote {\bibinfo {title} {Directing cell motions on micropatterned
  ratchets},}\ }\href@noop {} {\bibfield  {journal} {\bibinfo  {journal}
  {Nature physics}\ }\textbf {\bibinfo {volume} {5}},\ \bibinfo {pages} {606}
  (\bibinfo {year} {2009})}\BibitemShut {NoStop}%
\bibitem [{\citenamefont {Serreli}\ \emph {et~al.}(2007)\citenamefont
  {Serreli}, \citenamefont {Lee}, \citenamefont {Kay},\ and\ \citenamefont
  {Leigh}}]{serreli2007molecular}%
  \BibitemOpen
  \bibfield  {author} {\bibinfo {author} {\bibfnamefont {V.}~\bibnamefont
  {Serreli}}, \bibinfo {author} {\bibfnamefont {C.-F.}\ \bibnamefont {Lee}},
  \bibinfo {author} {\bibfnamefont {E.~R.}\ \bibnamefont {Kay}}, \ and\
  \bibinfo {author} {\bibfnamefont {D.~A.}\ \bibnamefont {Leigh}},\ }\bibfield
  {title} {\enquote {\bibinfo {title} {A molecular information ratchet},}\
  }\href@noop {} {\bibfield  {journal} {\bibinfo  {journal} {Nature}\ }\textbf
  {\bibinfo {volume} {445}},\ \bibinfo {pages} {523} (\bibinfo {year}
  {2007})}\BibitemShut {NoStop}%
\bibitem [{\citenamefont {Linke}\ \emph {et~al.}(1999)\citenamefont {Linke},
  \citenamefont {Humphrey}, \citenamefont {L{\"o}fgren}, \citenamefont
  {Sushkov}, \citenamefont {Newbury}, \citenamefont {Taylor},\ and\
  \citenamefont {Omling}}]{linke1999experimental}%
  \BibitemOpen
  \bibfield  {author} {\bibinfo {author} {\bibfnamefont {H.}~\bibnamefont
  {Linke}}, \bibinfo {author} {\bibfnamefont {T.}~\bibnamefont {Humphrey}},
  \bibinfo {author} {\bibfnamefont {A.}~\bibnamefont {L{\"o}fgren}}, \bibinfo
  {author} {\bibfnamefont {A.}~\bibnamefont {Sushkov}}, \bibinfo {author}
  {\bibfnamefont {R.}~\bibnamefont {Newbury}}, \bibinfo {author} {\bibfnamefont
  {R.}~\bibnamefont {Taylor}}, \ and\ \bibinfo {author} {\bibfnamefont
  {P.}~\bibnamefont {Omling}},\ }\bibfield  {title} {\enquote {\bibinfo {title}
  {Experimental tunneling ratchets},}\ }\href@noop {} {\bibfield  {journal}
  {\bibinfo  {journal} {Science}\ }\textbf {\bibinfo {volume} {286}},\ \bibinfo
  {pages} {2314} (\bibinfo {year} {1999})}\BibitemShut {NoStop}%
\bibitem [{\citenamefont {Costache}\ and\ \citenamefont
  {Valenzuela}(2010)}]{costache2010experimental}%
  \BibitemOpen
  \bibfield  {author} {\bibinfo {author} {\bibfnamefont {M.~V.}\ \bibnamefont
  {Costache}}\ and\ \bibinfo {author} {\bibfnamefont {S.~O.}\ \bibnamefont
  {Valenzuela}},\ }\bibfield  {title} {\enquote {\bibinfo {title} {Experimental
  spin ratchet},}\ }\href@noop {} {\bibfield  {journal} {\bibinfo  {journal}
  {Science}\ }\textbf {\bibinfo {volume} {330}},\ \bibinfo {pages} {1645}
  (\bibinfo {year} {2010})}\BibitemShut {NoStop}%
\bibitem [{\citenamefont {Drexler}\ \emph {et~al.}(2013)\citenamefont
  {Drexler}, \citenamefont {Tarasenko}, \citenamefont {Olbrich}, \citenamefont
  {Karch}, \citenamefont {Hirmer}, \citenamefont {M{\"u}ller}, \citenamefont
  {Gmitra}, \citenamefont {Fabian}, \citenamefont {Yakimova}, \citenamefont
  {Lara-Avila} \emph {et~al.}}]{drexler2013magnetic}%
  \BibitemOpen
  \bibfield  {author} {\bibinfo {author} {\bibfnamefont {C.}~\bibnamefont
  {Drexler}}, \bibinfo {author} {\bibfnamefont {S.}~\bibnamefont {Tarasenko}},
  \bibinfo {author} {\bibfnamefont {P.}~\bibnamefont {Olbrich}}, \bibinfo
  {author} {\bibfnamefont {J.}~\bibnamefont {Karch}}, \bibinfo {author}
  {\bibfnamefont {M.}~\bibnamefont {Hirmer}}, \bibinfo {author} {\bibfnamefont
  {F.}~\bibnamefont {M{\"u}ller}}, \bibinfo {author} {\bibfnamefont
  {M.}~\bibnamefont {Gmitra}}, \bibinfo {author} {\bibfnamefont
  {J.}~\bibnamefont {Fabian}}, \bibinfo {author} {\bibfnamefont
  {R.}~\bibnamefont {Yakimova}}, \bibinfo {author} {\bibfnamefont
  {S.}~\bibnamefont {Lara-Avila}},  \emph {et~al.},\ }\bibfield  {title}
  {\enquote {\bibinfo {title} {Magnetic quantum ratchet effect in graphene},}\
  }\href@noop {} {\bibfield  {journal} {\bibinfo  {journal} {Nature
  nanotechnology}\ }\textbf {\bibinfo {volume} {8}},\ \bibinfo {pages} {104}
  (\bibinfo {year} {2013})}\BibitemShut {NoStop}%
\bibitem [{\citenamefont {Lehmann}\ \emph {et~al.}(2002)\citenamefont
  {Lehmann}, \citenamefont {Kohler}, \citenamefont {H{\"a}nggi},\ and\
  \citenamefont {Nitzan}}]{lehmann2002molecular}%
  \BibitemOpen
  \bibfield  {author} {\bibinfo {author} {\bibfnamefont {J.}~\bibnamefont
  {Lehmann}}, \bibinfo {author} {\bibfnamefont {S.}~\bibnamefont {Kohler}},
  \bibinfo {author} {\bibfnamefont {P.}~\bibnamefont {H{\"a}nggi}}, \ and\
  \bibinfo {author} {\bibfnamefont {A.}~\bibnamefont {Nitzan}},\ }\bibfield
  {title} {\enquote {\bibinfo {title} {Molecular wires acting as coherent
  quantum ratchets},}\ }\href@noop {} {\bibfield  {journal} {\bibinfo
  {journal} {Physical review letters}\ }\textbf {\bibinfo {volume} {88}},\
  \bibinfo {pages} {228305} (\bibinfo {year} {2002})}\BibitemShut {NoStop}%
\bibitem [{\citenamefont {Zhang}\ \emph {et~al.}(2015)\citenamefont {Zhang},
  \citenamefont {Li},\ and\ \citenamefont {Guo}}]{zhang2015experimental}%
  \BibitemOpen
  \bibfield  {author} {\bibinfo {author} {\bibfnamefont {C.}~\bibnamefont
  {Zhang}}, \bibinfo {author} {\bibfnamefont {C.-F.}\ \bibnamefont {Li}}, \
  and\ \bibinfo {author} {\bibfnamefont {G.-C.}\ \bibnamefont {Guo}},\
  }\bibfield  {title} {\enquote {\bibinfo {title} {Experimental demonstration
  of photonic quantum ratchet},}\ }\href@noop {} {\bibfield  {journal}
  {\bibinfo  {journal} {Science Bulletin}\ }\textbf {\bibinfo {volume} {60}},\
  \bibinfo {pages} {249} (\bibinfo {year} {2015})}\BibitemShut {NoStop}%
\bibitem [{\citenamefont {Dreisow}\ \emph {et~al.}(2013)\citenamefont
  {Dreisow}, \citenamefont {Kartashov}, \citenamefont {Heinrich}, \citenamefont
  {Vysloukh}, \citenamefont {T{\"u}nnermann}, \citenamefont {Nolte},
  \citenamefont {Torner}, \citenamefont {Longhi},\ and\ \citenamefont
  {Szameit}}]{dreisow2013spatial}%
  \BibitemOpen
  \bibfield  {author} {\bibinfo {author} {\bibfnamefont {F.}~\bibnamefont
  {Dreisow}}, \bibinfo {author} {\bibfnamefont {Y.~V.}\ \bibnamefont
  {Kartashov}}, \bibinfo {author} {\bibfnamefont {M.}~\bibnamefont {Heinrich}},
  \bibinfo {author} {\bibfnamefont {V.~A.}\ \bibnamefont {Vysloukh}}, \bibinfo
  {author} {\bibfnamefont {A.}~\bibnamefont {T{\"u}nnermann}}, \bibinfo
  {author} {\bibfnamefont {S.}~\bibnamefont {Nolte}}, \bibinfo {author}
  {\bibfnamefont {L.}~\bibnamefont {Torner}}, \bibinfo {author} {\bibfnamefont
  {S.}~\bibnamefont {Longhi}}, \ and\ \bibinfo {author} {\bibfnamefont
  {A.}~\bibnamefont {Szameit}},\ }\bibfield  {title} {\enquote {\bibinfo
  {title} {Spatial light rectification in an optical waveguide lattice},}\
  }\href@noop {} {\bibfield  {journal} {\bibinfo  {journal} {EPL (Europhysics
  Letters)}\ }\textbf {\bibinfo {volume} {101}},\ \bibinfo {pages} {44002}
  (\bibinfo {year} {2013})}\BibitemShut {NoStop}%
\bibitem [{\citenamefont {Salger}\ \emph {et~al.}(2009)\citenamefont {Salger},
  \citenamefont {Kling}, \citenamefont {Hecking}, \citenamefont {Geckeler},
  \citenamefont {Morales-Molina},\ and\ \citenamefont
  {Weitz}}]{salger2009directed}%
  \BibitemOpen
  \bibfield  {author} {\bibinfo {author} {\bibfnamefont {T.}~\bibnamefont
  {Salger}}, \bibinfo {author} {\bibfnamefont {S.}~\bibnamefont {Kling}},
  \bibinfo {author} {\bibfnamefont {T.}~\bibnamefont {Hecking}}, \bibinfo
  {author} {\bibfnamefont {C.}~\bibnamefont {Geckeler}}, \bibinfo {author}
  {\bibfnamefont {L.}~\bibnamefont {Morales-Molina}}, \ and\ \bibinfo {author}
  {\bibfnamefont {M.}~\bibnamefont {Weitz}},\ }\bibfield  {title} {\enquote
  {\bibinfo {title} {Directed transport of atoms in a hamiltonian quantum
  ratchet},}\ }\href@noop {} {\bibfield  {journal} {\bibinfo  {journal}
  {Science}\ }\textbf {\bibinfo {volume} {326}},\ \bibinfo {pages} {1241}
  (\bibinfo {year} {2009})}\BibitemShut {NoStop}%
\bibitem [{\citenamefont {Ni}\ \emph {et~al.}(2017)\citenamefont {Ni},
  \citenamefont {Dadras}, \citenamefont {Lam}, \citenamefont {Shrestha},
  \citenamefont {Sadgrove}, \citenamefont {Wimberger},\ and\ \citenamefont
  {Summy}}]{ni2017hamiltonian}%
  \BibitemOpen
  \bibfield  {author} {\bibinfo {author} {\bibfnamefont {J.}~\bibnamefont
  {Ni}}, \bibinfo {author} {\bibfnamefont {S.}~\bibnamefont {Dadras}}, \bibinfo
  {author} {\bibfnamefont {W.~K.}\ \bibnamefont {Lam}}, \bibinfo {author}
  {\bibfnamefont {R.~K.}\ \bibnamefont {Shrestha}}, \bibinfo {author}
  {\bibfnamefont {M.}~\bibnamefont {Sadgrove}}, \bibinfo {author}
  {\bibfnamefont {S.}~\bibnamefont {Wimberger}}, \ and\ \bibinfo {author}
  {\bibfnamefont {G.~S.}\ \bibnamefont {Summy}},\ }\bibfield  {title} {\enquote
  {\bibinfo {title} {Hamiltonian ratchets with ultra-cold atoms},}\ }\href@noop
  {} {\bibfield  {journal} {\bibinfo  {journal} {Annalen der Physik}\ }\textbf
  {\bibinfo {volume} {529}},\ \bibinfo {pages} {1600335} (\bibinfo {year}
  {2017})}\BibitemShut {NoStop}%
\bibitem [{\citenamefont {H{\"a}nggi}\ and\ \citenamefont
  {Marchesoni}(2009)}]{hanggi2009artificial}%
  \BibitemOpen
  \bibfield  {author} {\bibinfo {author} {\bibfnamefont {P.}~\bibnamefont
  {H{\"a}nggi}}\ and\ \bibinfo {author} {\bibfnamefont {F.}~\bibnamefont
  {Marchesoni}},\ }\bibfield  {title} {\enquote {\bibinfo {title} {Artificial
  brownian motors: Controlling transport on the nanoscale},}\ }\href@noop {}
  {\bibfield  {journal} {\bibinfo  {journal} {Reviews of Modern Physics}\
  }\textbf {\bibinfo {volume} {81}},\ \bibinfo {pages} {387} (\bibinfo {year}
  {2009})}\BibitemShut {NoStop}%
\bibitem [{\citenamefont {Thouless}(1983)}]{thouless1983quantization}%
  \BibitemOpen
  \bibfield  {author} {\bibinfo {author} {\bibfnamefont {D.}~\bibnamefont
  {Thouless}},\ }\bibfield  {title} {\enquote {\bibinfo {title} {Quantization
  of particle transport},}\ }\href@noop {} {\bibfield  {journal} {\bibinfo
  {journal} {Physical Review B}\ }\textbf {\bibinfo {volume} {27}},\ \bibinfo
  {pages} {6083} (\bibinfo {year} {1983})}\BibitemShut {NoStop}%
\bibitem [{\citenamefont {Denisov}\ \emph {et~al.}(2007)\citenamefont
  {Denisov}, \citenamefont {Morales-Molina}, \citenamefont {Flach},\ and\
  \citenamefont {H{\"a}nggi}}]{denisov2007periodically}%
  \BibitemOpen
  \bibfield  {author} {\bibinfo {author} {\bibfnamefont {S.}~\bibnamefont
  {Denisov}}, \bibinfo {author} {\bibfnamefont {L.}~\bibnamefont
  {Morales-Molina}}, \bibinfo {author} {\bibfnamefont {S.}~\bibnamefont
  {Flach}}, \ and\ \bibinfo {author} {\bibfnamefont {P.}~\bibnamefont
  {H{\"a}nggi}},\ }\bibfield  {title} {\enquote {\bibinfo {title} {Periodically
  driven quantum ratchets: Symmetries and resonances},}\ }\href@noop {}
  {\bibfield  {journal} {\bibinfo  {journal} {Physical Review A}\ }\textbf
  {\bibinfo {volume} {75}},\ \bibinfo {pages} {063424} (\bibinfo {year}
  {2007})}\BibitemShut {NoStop}%
\bibitem [{\citenamefont {Gong}\ \emph {et~al.}(2007)\citenamefont {Gong},
  \citenamefont {Poletti},\ and\ \citenamefont
  {Hanggi}}]{gong2007dissipationless}%
  \BibitemOpen
  \bibfield  {author} {\bibinfo {author} {\bibfnamefont {J.}~\bibnamefont
  {Gong}}, \bibinfo {author} {\bibfnamefont {D.}~\bibnamefont {Poletti}}, \
  and\ \bibinfo {author} {\bibfnamefont {P.}~\bibnamefont {Hanggi}},\
  }\bibfield  {title} {\enquote {\bibinfo {title} {Dissipationless directed
  transport in rocked single-band quantum dynamics},}\ }\href@noop {}
  {\bibfield  {journal} {\bibinfo  {journal} {Physical Review A}\ }\textbf
  {\bibinfo {volume} {75}},\ \bibinfo {pages} {033602} (\bibinfo {year}
  {2007})}\BibitemShut {NoStop}%
\bibitem [{\citenamefont {Fedorova}\ \emph {et~al.}(2020)\citenamefont
  {Fedorova}, \citenamefont {Qiu}, \citenamefont {Linden},\ and\ \citenamefont
  {Kroha}}]{fedorova2020observation}%
  \BibitemOpen
  \bibfield  {author} {\bibinfo {author} {\bibfnamefont {Z.}~\bibnamefont
  {Fedorova}}, \bibinfo {author} {\bibfnamefont {H.}~\bibnamefont {Qiu}},
  \bibinfo {author} {\bibfnamefont {S.}~\bibnamefont {Linden}}, \ and\ \bibinfo
  {author} {\bibfnamefont {J.}~\bibnamefont {Kroha}},\ }\bibfield  {title}
  {\enquote {\bibinfo {title} {Observation of topological transport
  quantization by dissipation in fast thouless pumps},}\ }\href@noop {}
  {\bibfield  {journal} {\bibinfo  {journal} {Nature Communications}\ }\textbf
  {\bibinfo {volume} {11}},\ \bibinfo {pages} {1} (\bibinfo {year}
  {2020})}\BibitemShut {NoStop}%
\bibitem [{\citenamefont {Lohse}\ \emph {et~al.}(2016)\citenamefont {Lohse},
  \citenamefont {Schweizer}, \citenamefont {Zilberberg}, \citenamefont
  {Aidelsburger},\ and\ \citenamefont {Bloch}}]{lohse2016thouless}%
  \BibitemOpen
  \bibfield  {author} {\bibinfo {author} {\bibfnamefont {M.}~\bibnamefont
  {Lohse}}, \bibinfo {author} {\bibfnamefont {C.}~\bibnamefont {Schweizer}},
  \bibinfo {author} {\bibfnamefont {O.}~\bibnamefont {Zilberberg}}, \bibinfo
  {author} {\bibfnamefont {M.}~\bibnamefont {Aidelsburger}}, \ and\ \bibinfo
  {author} {\bibfnamefont {I.}~\bibnamefont {Bloch}},\ }\bibfield  {title}
  {\enquote {\bibinfo {title} {A thouless quantum pump with ultracold bosonic
  atoms in an optical superlattice},}\ }\href@noop {} {\bibfield  {journal}
  {\bibinfo  {journal} {Nature Physics}\ }\textbf {\bibinfo {volume} {12}},\
  \bibinfo {pages} {350} (\bibinfo {year} {2016})}\BibitemShut {NoStop}%
\bibitem [{\citenamefont {Nakajima}\ \emph {et~al.}(2016)\citenamefont
  {Nakajima}, \citenamefont {Tomita}, \citenamefont {Taie}, \citenamefont
  {Ichinose}, \citenamefont {Ozawa}, \citenamefont {Wang}, \citenamefont
  {Troyer},\ and\ \citenamefont {Takahashi}}]{nakajima2016topological}%
  \BibitemOpen
  \bibfield  {author} {\bibinfo {author} {\bibfnamefont {S.}~\bibnamefont
  {Nakajima}}, \bibinfo {author} {\bibfnamefont {T.}~\bibnamefont {Tomita}},
  \bibinfo {author} {\bibfnamefont {S.}~\bibnamefont {Taie}}, \bibinfo {author}
  {\bibfnamefont {T.}~\bibnamefont {Ichinose}}, \bibinfo {author}
  {\bibfnamefont {H.}~\bibnamefont {Ozawa}}, \bibinfo {author} {\bibfnamefont
  {L.}~\bibnamefont {Wang}}, \bibinfo {author} {\bibfnamefont {M.}~\bibnamefont
  {Troyer}}, \ and\ \bibinfo {author} {\bibfnamefont {Y.}~\bibnamefont
  {Takahashi}},\ }\bibfield  {title} {\enquote {\bibinfo {title} {Topological
  thouless pumping of ultracold fermions},}\ }\href@noop {} {\bibfield
  {journal} {\bibinfo  {journal} {Nature Physics}\ }\textbf {\bibinfo {volume}
  {12}},\ \bibinfo {pages} {296} (\bibinfo {year} {2016})}\BibitemShut
  {NoStop}%
\bibitem [{\citenamefont {H\"ockendorf}\ \emph {et~al.}(2020)\citenamefont
  {H\"ockendorf}, \citenamefont {Alvermann},\ and\ \citenamefont
  {Fehske}}]{fehske2020quantized}%
  \BibitemOpen
  \bibfield  {author} {\bibinfo {author} {\bibfnamefont {B.}~\bibnamefont
  {H\"ockendorf}}, \bibinfo {author} {\bibfnamefont {A.}~\bibnamefont
  {Alvermann}}, \ and\ \bibinfo {author} {\bibfnamefont {H.}~\bibnamefont
  {Fehske}},\ }\bibfield  {title} {\enquote {\bibinfo {title} {Topological
  origin of quantized transport in non-hermitian floquet chains},}\ }\href
  {http://dx.doi.org/10.1103/PhysRevResearch.2.023235} {\bibfield  {journal}
  {\bibinfo  {journal} {Phys. Rev. Research}\ }\textbf {\bibinfo {volume}
  {2}},\ \bibinfo {pages} {023235} (\bibinfo {year} {2020})}\BibitemShut
  {NoStop}%
\bibitem [{\citenamefont {Su}\ \emph {et~al.}(1979)\citenamefont {Su},
  \citenamefont {Schrieffer},\ and\ \citenamefont {Heeger}}]{su1979solitons}%
  \BibitemOpen
  \bibfield  {author} {\bibinfo {author} {\bibfnamefont {W.}~\bibnamefont
  {Su}}, \bibinfo {author} {\bibfnamefont {J.}~\bibnamefont {Schrieffer}}, \
  and\ \bibinfo {author} {\bibfnamefont {A.~J.}\ \bibnamefont {Heeger}},\
  }\bibfield  {title} {\enquote {\bibinfo {title} {Solitons in
  polyacetylene},}\ }\href@noop {} {\bibfield  {journal} {\bibinfo  {journal}
  {Physical review letters}\ }\textbf {\bibinfo {volume} {42}},\ \bibinfo
  {pages} {1698} (\bibinfo {year} {1979})}\BibitemShut {NoStop}%
\bibitem [{\citenamefont {Kartashov}\ \emph {et~al.}(2016)\citenamefont
  {Kartashov}, \citenamefont {Vysloukh}, \citenamefont {Konotop},\ and\
  \citenamefont {Torner}}]{kartashov2016diffraction}%
  \BibitemOpen
  \bibfield  {author} {\bibinfo {author} {\bibfnamefont {Y.~V.}\ \bibnamefont
  {Kartashov}}, \bibinfo {author} {\bibfnamefont {V.~A.}\ \bibnamefont
  {Vysloukh}}, \bibinfo {author} {\bibfnamefont {V.~V.}\ \bibnamefont
  {Konotop}}, \ and\ \bibinfo {author} {\bibfnamefont {L.}~\bibnamefont
  {Torner}},\ }\bibfield  {title} {\enquote {\bibinfo {title} {Diffraction
  control in p t-symmetric photonic lattices: From beam rectification to
  dynamic localization},}\ }\href@noop {} {\bibfield  {journal} {\bibinfo
  {journal} {Physical Review A}\ }\textbf {\bibinfo {volume} {93}},\ \bibinfo
  {pages} {013841} (\bibinfo {year} {2016})}\BibitemShut {NoStop}%
\bibitem [{\citenamefont {Budich}\ \emph {et~al.}(2017)\citenamefont {Budich},
  \citenamefont {Hu},\ and\ \citenamefont {Zoller}}]{budich2017helical}%
  \BibitemOpen
  \bibfield  {author} {\bibinfo {author} {\bibfnamefont {J.~C.}\ \bibnamefont
  {Budich}}, \bibinfo {author} {\bibfnamefont {Y.}~\bibnamefont {Hu}}, \ and\
  \bibinfo {author} {\bibfnamefont {P.}~\bibnamefont {Zoller}},\ }\bibfield
  {title} {\enquote {\bibinfo {title} {Helical floquet channels in 1d
  lattices},}\ }\href@noop {} {\bibfield  {journal} {\bibinfo  {journal}
  {Physical review letters}\ }\textbf {\bibinfo {volume} {118}},\ \bibinfo
  {pages} {105302} (\bibinfo {year} {2017})}\BibitemShut {NoStop}%
\bibitem [{\citenamefont {Thuberg}\ \emph {et~al.}(2017)\citenamefont
  {Thuberg}, \citenamefont {Mu{\~n}oz}, \citenamefont {Eggert},\ and\
  \citenamefont {Reyes}}]{thuberg2017perfect}%
  \BibitemOpen
  \bibfield  {author} {\bibinfo {author} {\bibfnamefont {D.}~\bibnamefont
  {Thuberg}}, \bibinfo {author} {\bibfnamefont {E.}~\bibnamefont {Mu{\~n}oz}},
  \bibinfo {author} {\bibfnamefont {S.}~\bibnamefont {Eggert}}, \ and\ \bibinfo
  {author} {\bibfnamefont {S.~A.}\ \bibnamefont {Reyes}},\ }\bibfield  {title}
  {\enquote {\bibinfo {title} {Perfect spin filter by periodic drive of a
  ferromagnetic quantum barrier},}\ }\href@noop {} {\bibfield  {journal}
  {\bibinfo  {journal} {Physical Review Letters}\ }\textbf {\bibinfo {volume}
  {119}},\ \bibinfo {pages} {267701} (\bibinfo {year} {2017})}\BibitemShut
  {NoStop}%
\bibitem [{\citenamefont {Reyes}\ \emph {et~al.}(2017)\citenamefont {Reyes},
  \citenamefont {Thuberg}, \citenamefont {P{\'e}rez}, \citenamefont {Dauer},\
  and\ \citenamefont {Eggert}}]{reyes2017transport}%
  \BibitemOpen
  \bibfield  {author} {\bibinfo {author} {\bibfnamefont {S.~A.}\ \bibnamefont
  {Reyes}}, \bibinfo {author} {\bibfnamefont {D.}~\bibnamefont {Thuberg}},
  \bibinfo {author} {\bibfnamefont {D.}~\bibnamefont {P{\'e}rez}}, \bibinfo
  {author} {\bibfnamefont {C.}~\bibnamefont {Dauer}}, \ and\ \bibinfo {author}
  {\bibfnamefont {S.}~\bibnamefont {Eggert}},\ }\bibfield  {title} {\enquote
  {\bibinfo {title} {Transport through an ac-driven impurity: Fano interference
  and bound states in the continuum},}\ }\href@noop {} {\bibfield  {journal}
  {\bibinfo  {journal} {New Journal of Physics}\ }\textbf {\bibinfo {volume}
  {19}},\ \bibinfo {pages} {043029} (\bibinfo {year} {2017})}\BibitemShut
  {NoStop}%
\bibitem [{\citenamefont {Agarwala}\ and\ \citenamefont
  {Sen}(2017)}]{agarwala2017effects}%
  \BibitemOpen
  \bibfield  {author} {\bibinfo {author} {\bibfnamefont {A.}~\bibnamefont
  {Agarwala}}\ and\ \bibinfo {author} {\bibfnamefont {D.}~\bibnamefont {Sen}},\
  }\bibfield  {title} {\enquote {\bibinfo {title} {Effects of local periodic
  driving on transport and generation of bound states},}\ }\href@noop {}
  {\bibfield  {journal} {\bibinfo  {journal} {Physical Review B}\ }\textbf
  {\bibinfo {volume} {96}},\ \bibinfo {pages} {104309} (\bibinfo {year}
  {2017})}\BibitemShut {NoStop}%
\bibitem [{\citenamefont {Moskalets}\ and\ \citenamefont
  {B\"uttiker}(2002)}]{Moskalets2002}%
  \BibitemOpen
  \bibfield  {author} {\bibinfo {author} {\bibfnamefont {M.}~\bibnamefont
  {Moskalets}}\ and\ \bibinfo {author} {\bibfnamefont {M.}~\bibnamefont
  {B\"uttiker}},\ }\bibfield  {title} {\enquote {\bibinfo {title} {Floquet
  scattering theory of quantum pumps},}\ }\href
  {http://dx.doi.org/10.1103/PhysRevB.66.205320} {\bibfield  {journal}
  {\bibinfo  {journal} {Phys. Rev. B}\ }\textbf {\bibinfo {volume} {66}},\
  \bibinfo {pages} {205320} (\bibinfo {year} {2002})}\BibitemShut {NoStop}%
\bibitem [{\citenamefont {Li}\ \emph {et~al.}(2018{\natexlab{a}})\citenamefont
  {Li}, \citenamefont {Shapiro},\ and\ \citenamefont {Kottos}}]{Li2018}%
  \BibitemOpen
  \bibfield  {author} {\bibinfo {author} {\bibfnamefont {H.}~\bibnamefont
  {Li}}, \bibinfo {author} {\bibfnamefont {B.}~\bibnamefont {Shapiro}}, \ and\
  \bibinfo {author} {\bibfnamefont {T.}~\bibnamefont {Kottos}},\ }\bibfield
  {title} {\enquote {\bibinfo {title} {Floquet scattering theory based on
  effective hamiltonians of driven systems},}\ }\href
  {http://dx.doi.org/10.1103/PhysRevB.98.121101} {\bibfield  {journal}
  {\bibinfo  {journal} {Phys. Rev. B}\ }\textbf {\bibinfo {volume} {98}},\
  \bibinfo {pages} {121101} (\bibinfo {year} {2018}{\natexlab{a}})}\BibitemShut
  {NoStop}%
\bibitem [{\citenamefont {Smith}(2015)}]{Smith2015}%
  \BibitemOpen
  \bibfield  {author} {\bibinfo {author} {\bibfnamefont {D.~H.}\ \bibnamefont
  {Smith}},\ }\bibfield  {title} {\enquote {\bibinfo {title} {{Inducing
  Resonant Interactions in Ultracold Atoms with a Modulated Magnetic Field}},}\
  }\href {http://dx.doi.org/10.1103/PhysRevLett.115.193002} {\bibfield
  {journal} {\bibinfo  {journal} {Phys. Rev. Lett}\ }\textbf {\bibinfo {volume}
  {115}},\ \bibinfo {pages} {193002} (\bibinfo {year} {2015})}\BibitemShut
  {NoStop}%
\bibitem [{\citenamefont {Millack}(1990)}]{Millack1990}%
  \BibitemOpen
  \bibfield  {author} {\bibinfo {author} {\bibfnamefont {T.}~\bibnamefont
  {Millack}},\ }\bibfield  {title} {\enquote {\bibinfo {title} {T-matrix and
  k-matrix floquet theory for atoms in strong laser fields},}\ }\href
  {http://dx.doi.org/10.1088/0953-4075/23/11/009} {\bibfield  {journal}
  {\bibinfo  {journal} {Journal of Physics B: Atomic, Molecular and Optical
  Physics}\ }\textbf {\bibinfo {volume} {23}},\ \bibinfo {pages} {1693}
  (\bibinfo {year} {1990})}\BibitemShut {NoStop}%
\bibitem [{\citenamefont {G{\'o}mez-Le{\'o}n}\ and\ \citenamefont
  {Platero}(2013)}]{gomez2013floquet}%
  \BibitemOpen
  \bibfield  {author} {\bibinfo {author} {\bibfnamefont {A.}~\bibnamefont
  {G{\'o}mez-Le{\'o}n}}\ and\ \bibinfo {author} {\bibfnamefont
  {G.}~\bibnamefont {Platero}},\ }\bibfield  {title} {\enquote {\bibinfo
  {title} {Floquet-bloch theory and topology in periodically driven
  lattices},}\ }\href@noop {} {\bibfield  {journal} {\bibinfo  {journal}
  {Physical review letters}\ }\textbf {\bibinfo {volume} {110}},\ \bibinfo
  {pages} {200403} (\bibinfo {year} {2013})}\BibitemShut {NoStop}%
\bibitem [{\citenamefont {Garanovich}\ \emph {et~al.}(2012)\citenamefont
  {Garanovich}, \citenamefont {Longhi}, \citenamefont {Sukhorukov},\ and\
  \citenamefont {Kivshar}}]{garanovich2012light}%
  \BibitemOpen
  \bibfield  {author} {\bibinfo {author} {\bibfnamefont {I.~L.}\ \bibnamefont
  {Garanovich}}, \bibinfo {author} {\bibfnamefont {S.}~\bibnamefont {Longhi}},
  \bibinfo {author} {\bibfnamefont {A.~A.}\ \bibnamefont {Sukhorukov}}, \ and\
  \bibinfo {author} {\bibfnamefont {Y.~S.}\ \bibnamefont {Kivshar}},\
  }\bibfield  {title} {\enquote {\bibinfo {title} {Light propagation and
  localization in modulated photonic lattices and waveguides},}\ }\href@noop {}
  {\bibfield  {journal} {\bibinfo  {journal} {Physics Reports}\ }\textbf
  {\bibinfo {volume} {518}},\ \bibinfo {pages} {1} (\bibinfo {year}
  {2012})}\BibitemShut {NoStop}%
\bibitem [{\citenamefont {Bleckmann}\ \emph {et~al.}(2017)\citenamefont
  {Bleckmann}, \citenamefont {Cherpakova}, \citenamefont {Linden},\ and\
  \citenamefont {Alberti}}]{bleckmann2017spectral}%
  \BibitemOpen
  \bibfield  {author} {\bibinfo {author} {\bibfnamefont {F.}~\bibnamefont
  {Bleckmann}}, \bibinfo {author} {\bibfnamefont {Z.}~\bibnamefont
  {Cherpakova}}, \bibinfo {author} {\bibfnamefont {S.}~\bibnamefont {Linden}},
  \ and\ \bibinfo {author} {\bibfnamefont {A.}~\bibnamefont {Alberti}},\
  }\bibfield  {title} {\enquote {\bibinfo {title} {Spectral imaging of
  topological edge states in plasmonic waveguide arrays},}\ }\href@noop {}
  {\bibfield  {journal} {\bibinfo  {journal} {Physical Review B}\ }\textbf
  {\bibinfo {volume} {96}},\ \bibinfo {pages} {045417} (\bibinfo {year}
  {2017})}\BibitemShut {NoStop}%
\bibitem [{\citenamefont {Cherpakova}\ \emph {et~al.}(2017)\citenamefont
  {Cherpakova}, \citenamefont {Bleckmann}, \citenamefont {Vogler},\ and\
  \citenamefont {Linden}}]{cherpakova2017transverse}%
  \BibitemOpen
  \bibfield  {author} {\bibinfo {author} {\bibfnamefont {Z.}~\bibnamefont
  {Cherpakova}}, \bibinfo {author} {\bibfnamefont {F.}~\bibnamefont
  {Bleckmann}}, \bibinfo {author} {\bibfnamefont {T.}~\bibnamefont {Vogler}}, \
  and\ \bibinfo {author} {\bibfnamefont {S.}~\bibnamefont {Linden}},\
  }\bibfield  {title} {\enquote {\bibinfo {title} {Transverse anderson
  localization of surface plasmon polaritons},}\ }\href@noop {} {\bibfield
  {journal} {\bibinfo  {journal} {Optics letters}\ }\textbf {\bibinfo {volume}
  {42}},\ \bibinfo {pages} {2165} (\bibinfo {year} {2017})}\BibitemShut
  {NoStop}%
\bibitem [{\citenamefont {Rechtsman}\ \emph {et~al.}(2013)\citenamefont
  {Rechtsman}, \citenamefont {Zeuner}, \citenamefont {Plotnik}, \citenamefont
  {Lumer}, \citenamefont {Podolsky}, \citenamefont {Dreisow}, \citenamefont
  {Nolte}, \citenamefont {Segev},\ and\ \citenamefont
  {Szameit}}]{rechtsman2013photonic}%
  \BibitemOpen
  \bibfield  {author} {\bibinfo {author} {\bibfnamefont {M.~C.}\ \bibnamefont
  {Rechtsman}}, \bibinfo {author} {\bibfnamefont {J.~M.}\ \bibnamefont
  {Zeuner}}, \bibinfo {author} {\bibfnamefont {Y.}~\bibnamefont {Plotnik}},
  \bibinfo {author} {\bibfnamefont {Y.}~\bibnamefont {Lumer}}, \bibinfo
  {author} {\bibfnamefont {D.}~\bibnamefont {Podolsky}}, \bibinfo {author}
  {\bibfnamefont {F.}~\bibnamefont {Dreisow}}, \bibinfo {author} {\bibfnamefont
  {S.}~\bibnamefont {Nolte}}, \bibinfo {author} {\bibfnamefont
  {M.}~\bibnamefont {Segev}}, \ and\ \bibinfo {author} {\bibfnamefont
  {A.}~\bibnamefont {Szameit}},\ }\bibfield  {title} {\enquote {\bibinfo
  {title} {Photonic floquet topological insulators},}\ }\href@noop {}
  {\bibfield  {journal} {\bibinfo  {journal} {Nature}\ }\textbf {\bibinfo
  {volume} {496}},\ \bibinfo {pages} {196} (\bibinfo {year}
  {2013})}\BibitemShut {NoStop}%
\bibitem [{\citenamefont {Fedorova}\ \emph {et~al.}(2019)\citenamefont
  {Fedorova}, \citenamefont {J{\"o}rg}, \citenamefont {Dauer}, \citenamefont
  {Letscher}, \citenamefont {Fleischhauer}, \citenamefont {Eggert},
  \citenamefont {Linden},\ and\ \citenamefont {von
  Freymann}}]{fedorova2019limits}%
  \BibitemOpen
  \bibfield  {author} {\bibinfo {author} {\bibfnamefont {Z.}~\bibnamefont
  {Fedorova}}, \bibinfo {author} {\bibfnamefont {C.}~\bibnamefont {J{\"o}rg}},
  \bibinfo {author} {\bibfnamefont {C.}~\bibnamefont {Dauer}}, \bibinfo
  {author} {\bibfnamefont {F.}~\bibnamefont {Letscher}}, \bibinfo {author}
  {\bibfnamefont {M.}~\bibnamefont {Fleischhauer}}, \bibinfo {author}
  {\bibfnamefont {S.}~\bibnamefont {Eggert}}, \bibinfo {author} {\bibfnamefont
  {S.}~\bibnamefont {Linden}}, \ and\ \bibinfo {author} {\bibfnamefont
  {G.}~\bibnamefont {von Freymann}},\ }\bibfield  {title} {\enquote {\bibinfo
  {title} {Limits of topological protection under local periodic driving},}\
  }\href@noop {} {\bibfield  {journal} {\bibinfo  {journal} {Light: Science \&
  Applications}\ }\textbf {\bibinfo {volume} {8}},\ \bibinfo {pages} {1}
  (\bibinfo {year} {2019})}\BibitemShut {NoStop}%
\bibitem [{\citenamefont {Guo}\ \emph {et~al.}(2009)\citenamefont {Guo},
  \citenamefont {Salamo}, \citenamefont {Duchesne}, \citenamefont {Morandotti},
  \citenamefont {Volatier-Ravat}, \citenamefont {Aimez}, \citenamefont
  {Siviloglou},\ and\ \citenamefont {Christodoulides}}]{guo2009observation}%
  \BibitemOpen
  \bibfield  {author} {\bibinfo {author} {\bibfnamefont {A.}~\bibnamefont
  {Guo}}, \bibinfo {author} {\bibfnamefont {G.}~\bibnamefont {Salamo}},
  \bibinfo {author} {\bibfnamefont {D.}~\bibnamefont {Duchesne}}, \bibinfo
  {author} {\bibfnamefont {R.}~\bibnamefont {Morandotti}}, \bibinfo {author}
  {\bibfnamefont {M.}~\bibnamefont {Volatier-Ravat}}, \bibinfo {author}
  {\bibfnamefont {V.}~\bibnamefont {Aimez}}, \bibinfo {author} {\bibfnamefont
  {G.}~\bibnamefont {Siviloglou}}, \ and\ \bibinfo {author} {\bibfnamefont
  {D.}~\bibnamefont {Christodoulides}},\ }\bibfield  {title} {\enquote
  {\bibinfo {title} {Observation of p t-symmetry breaking in complex optical
  potentials},}\ }\href@noop {} {\bibfield  {journal} {\bibinfo  {journal}
  {Physical Review Letters}\ }\textbf {\bibinfo {volume} {103}},\ \bibinfo
  {pages} {093902} (\bibinfo {year} {2009})}\BibitemShut {NoStop}%
\bibitem [{\citenamefont {Song}\ \emph {et~al.}(2019)\citenamefont {Song},
  \citenamefont {Sun}, \citenamefont {Chen}, \citenamefont {Song},
  \citenamefont {Xiao}, \citenamefont {Zhu},\ and\ \citenamefont
  {Li}}]{song2019breakup}%
  \BibitemOpen
  \bibfield  {author} {\bibinfo {author} {\bibfnamefont {W.}~\bibnamefont
  {Song}}, \bibinfo {author} {\bibfnamefont {W.}~\bibnamefont {Sun}}, \bibinfo
  {author} {\bibfnamefont {C.}~\bibnamefont {Chen}}, \bibinfo {author}
  {\bibfnamefont {Q.}~\bibnamefont {Song}}, \bibinfo {author} {\bibfnamefont
  {S.}~\bibnamefont {Xiao}}, \bibinfo {author} {\bibfnamefont {S.}~\bibnamefont
  {Zhu}}, \ and\ \bibinfo {author} {\bibfnamefont {T.}~\bibnamefont {Li}},\
  }\bibfield  {title} {\enquote {\bibinfo {title} {Breakup and recovery of
  topological zero modes in finite non-hermitian optical lattices},}\
  }\href@noop {} {\bibfield  {journal} {\bibinfo  {journal} {Physical review
  letters}\ }\textbf {\bibinfo {volume} {123}},\ \bibinfo {pages} {165701}
  (\bibinfo {year} {2019})}\BibitemShut {NoStop}%
\bibitem [{\citenamefont {Longhi}\ \emph {et~al.}(2015)\citenamefont {Longhi},
  \citenamefont {Gatti},\ and\ \citenamefont {Della~Valle}}]{Longhi2015}%
  \BibitemOpen
  \bibfield  {author} {\bibinfo {author} {\bibfnamefont {S.}~\bibnamefont
  {Longhi}}, \bibinfo {author} {\bibfnamefont {D.}~\bibnamefont {Gatti}}, \
  and\ \bibinfo {author} {\bibfnamefont {G.}~\bibnamefont {Della~Valle}},\
  }\bibfield  {title} {\enquote {\bibinfo {title} {Non-hermitian transparency
  and one-way transport in low-dimensional lattices by an imaginary gauge
  field},}\ }\href {http://dx.doi.org/10.1103/PhysRevB.92.094204} {\bibfield
  {journal} {\bibinfo  {journal} {Phys. Rev. B}\ }\textbf {\bibinfo {volume}
  {92}},\ \bibinfo {pages} {094204} (\bibinfo {year} {2015})}\BibitemShut
  {NoStop}%
\bibitem [{\citenamefont {Eckardt}(2017)}]{Eckardt2017}%
  \BibitemOpen
  \bibfield  {author} {\bibinfo {author} {\bibfnamefont {A.}~\bibnamefont
  {Eckardt}},\ }\bibfield  {title} {\enquote {\bibinfo {title} {{Colloquium:
  Atomic quantum gases in periodically driven optical lattices}},}\ }\href
  {http://dx.doi.org/10.1103/RevModPhys.89.011004} {\bibfield  {journal}
  {\bibinfo  {journal} {Rev. Mod. Phys.}\ }\textbf {\bibinfo {volume} {89}},\
  \bibinfo {pages} {1} (\bibinfo {year} {2017})}\BibitemShut {NoStop}%
\bibitem [{\citenamefont {Sambe}(1973)}]{Sambe1973}%
  \BibitemOpen
  \bibfield  {author} {\bibinfo {author} {\bibfnamefont {H.}~\bibnamefont
  {Sambe}},\ }\bibfield  {title} {\enquote {\bibinfo {title} {{Steady states
  and quasienergies of a quantum-mechanical system in an oscillating field}},}\
  }\href {http://dx.doi.org/10.1103/PhysRevA.7.2203} {\bibfield  {journal}
  {\bibinfo  {journal} {Phys. Rev. A}\ }\textbf {\bibinfo {volume} {7}},\
  \bibinfo {pages} {2203} (\bibinfo {year} {1973})}\BibitemShut {NoStop}%
\bibitem [{\citenamefont {Li}\ \emph {et~al.}(2018{\natexlab{b}})\citenamefont
  {Li}, \citenamefont {Kottos},\ and\ \citenamefont {Shapiro}}]{Li2018a}%
  \BibitemOpen
  \bibfield  {author} {\bibinfo {author} {\bibfnamefont {H.}~\bibnamefont
  {Li}}, \bibinfo {author} {\bibfnamefont {T.}~\bibnamefont {Kottos}}, \ and\
  \bibinfo {author} {\bibfnamefont {B.}~\bibnamefont {Shapiro}},\ }\bibfield
  {title} {\enquote {\bibinfo {title} {Floquet-network theory of nonreciprocal
  transport},}\ }\href {http://dx.doi.org/10.1103/PhysRevApplied.9.044031}
  {\bibfield  {journal} {\bibinfo  {journal} {Phys. Rev. Applied}\ }\textbf
  {\bibinfo {volume} {9}},\ \bibinfo {pages} {044031} (\bibinfo {year}
  {2018}{\natexlab{b}})}\BibitemShut {NoStop}%
\bibitem [{\citenamefont {Block}\ \emph {et~al.}(2014)\citenamefont {Block},
  \citenamefont {Etrich}, \citenamefont {Limboeck}, \citenamefont {Bleckmann},
  \citenamefont {Soergel}, \citenamefont {Rockstuhl},\ and\ \citenamefont
  {Linden}}]{block2014bloch}%
  \BibitemOpen
  \bibfield  {author} {\bibinfo {author} {\bibfnamefont {A.}~\bibnamefont
  {Block}}, \bibinfo {author} {\bibfnamefont {C.}~\bibnamefont {Etrich}},
  \bibinfo {author} {\bibfnamefont {T.}~\bibnamefont {Limboeck}}, \bibinfo
  {author} {\bibfnamefont {F.}~\bibnamefont {Bleckmann}}, \bibinfo {author}
  {\bibfnamefont {E.}~\bibnamefont {Soergel}}, \bibinfo {author} {\bibfnamefont
  {C.}~\bibnamefont {Rockstuhl}}, \ and\ \bibinfo {author} {\bibfnamefont
  {S.}~\bibnamefont {Linden}},\ }\bibfield  {title} {\enquote {\bibinfo {title}
  {Bloch oscillations in plasmonic waveguide arrays},}\ }\href@noop {}
  {\bibfield  {journal} {\bibinfo  {journal} {Nature communications}\ }\textbf
  {\bibinfo {volume} {5}},\ \bibinfo {pages} {1} (\bibinfo {year}
  {2014})}\BibitemShut {NoStop}%
\end{thebibliography}%

\end{document}